\documentclass[submission, Phys]{SciPost}

\hypersetup{
    colorlinks,
    linkcolor={red!50!black},
    citecolor={blue!50!black},
    urlcolor={blue!80!black}
}

\usepackage[bitstream-charter]{mathdesign}
\DeclareSymbolFont{usualmathcal}{OMS}{cmsy}{m}{n}
\DeclareSymbolFontAlphabet{\mathcal}{usualmathcal}

\fancypagestyle{SPstyle}{
\fancyhf{}
\lhead{\colorbox{scipostblue}{\bf \color{white} ~SciPost Physics }}
\rhead{{\bf \color{scipostdeepblue} ~Submission }}

\fancyfoot[C]{\textbf{\thepage}}
}

\usepackage{amsmath,physics,dsfont}

\newcommand{\Jcal}{\mathcal{J}}
\newcommand{\Jbb}{\mathbb{J}}
\newcommand{\Kbb}{\mathbb{K}}

\newcommand{\Lcal}{\mathcal{L}}
\newcommand{\Dcal}{\mathcal{D}}
\newcommand{\Rcal}{\mathcal{R}}
\newcommand{\Kcal}{\mathcal{K}}
\newcommand{\Scal}{\mathcal{S}}

\newcommand{\T}{\scalebox{0.6}{\text{T}}}
\newcommand{\rD}{{}^{\scalebox{0.6}{(\textit{D})}}}
\newcommand{\rd}{{}^{\scalebox{0.6}{(\textit{d})}}}
\newcommand{\mpl}{m_{\mathrm{pl}}^{d-2}}
\newcommand{\dint}{\!\int\!\!\mathrm{d}^d\!}
\newcommand{\Dint}{\!\int\!\!\mathrm{d}^{\scalebox{0.6}{\textit{D}}}\!}
\newcommand{\id}{\mathds{1}}

\begin{document}

\pagestyle{SPstyle}

\begin{center}{\Large \textbf{\color{scipostdeepblue}{
Canonical Vielbeins for general relativity: \\$D+1$ Decomposition and Constraint Analysis
}}}\end{center}

\begin{center}\textbf{
Joakim Flinckman\textsuperscript{1$\star$} and
Daniel Blixt\textsuperscript{2$\dagger$}
}\end{center}

\begin{center}
{\bf 1} Department of Physics, Stockholm University, SE-106 91 Stockholm, Sweden
\\
{\bf 2} Department of Mathematical Sciences, Chalmers University of Technology and \newline University of Gothenburg, SE-412 96, Gothenburg, Sweden
\\[\baselineskip]
$\star$ \href{mailto:Joakim.Flinckman@fysik.su.se}{\small Joakim.Flinckman@fysik.su.se}\,,\quad
$\dagger$ \href{mailto:blixtd@chalmers.se}{\small blixtd@chalmers.se}
\end{center}

\section*{\color{scipostdeepblue}{Abstract}}
\textbf{\boldmath{We provide a self-contained derivation of the Hamiltonian formulation of general relativity in vielbein variables in $d=D+1$ dimensions. Starting from the Einstein--Hilbert action in a standard metric $D+1$ decomposition, we derive Lorentz- and $\mathrm{SO}(D)$-covariant phase-space actions, identify the primary Lorentz constraints, and obtain the Hamiltonian and momentum constraints. We compute the resulting first-class constraint algebra, relate the vielbein and metric phase-space formulations, and discuss the rotation/boost decomposition. In particular, we construct the boost generator in the $\mathrm{SO}(D)$-covariant formulation, thereby recovering full local Lorentz symmetry.}}

\vspace{\baselineskip}



\noindent\rule{\textwidth}{1pt}
\tableofcontents
\noindent\rule{\textwidth}{1pt}

\section{Introduction}
\label{sec:intro}
General relativity was originally formulated in terms of a metric field $g_{\mu \nu}(x)$, but it was recognised early--by Cartan \cite{Cartan:1923} and Einstein himself \cite{Einstein:1928}--that it is sometimes convenient or even necessary (e.g. when working with spinors \cite{Weyl:1929,Fock1929}), to formulate the theory in terms of vielbeins, also called frame fields, vierbeins, or tetrads in 3+1 dimensions. 

In other situations, it is convenient to perform a foliation of spacetime, singling out a time direction, and working with spatial hypersurfaces in terms of a $D{+}1$-decomposition. Spacetime decompositions have a long history and have been used for different purposes. For example: when deriving the Hamiltonian and constraint algebra \cite{Rosenfeld:1930, Bergmann:1949, BergmannBrunings:1949, BergmannPenfieldSchillerZatzkis:1950, PiraniSchild:1950, Dirac:1950, Dirac:1951, AndersonBergmann:1951, PiraniSchildSkinner:1952, Dirac:1958, Dirac:1959}, when studying the initial value problem \cite{Darmois:1927, Lichnerowicz:1944, FouresBruhat:1952}, or when performing numerical evolution of the field equations \cite{ShibataNakamura:1995, BaumgarteShapiro:1998, Pretorius:2005, CampanelliLoustoMarronettiZlochower:2006, BakerCentrellaChoiKoppitzVanMeter:2006}.

The Hamiltonian formulation played an important role in the early days of trying to quantise general relativity \cite{Schwinger:1963,DeWitt:1967yk}. Although Arnowitt, Deser and Misner were not the first to write down the modern $D{+}1$-decomposition,\footnote{The important work of Rosenfeld \cite{Rosenfeld:1940}, Bergmann and  Anderson \cite{Bergmann:1949zz,Anderson:1951ta}, and Dirac \cite{Dirac:1958sc, Dirac:1958jc} is often overlooked.} their work is among the most cited references, which has led to the widespread use of the name ``ADM decomposition´´ and ``ADM variables´´. However, we will use the more author-neutral $D{+}1$-decomposition.

An early treatment of spacetime decomposition of the vielbein form of the Einstein–Hilbert action was presented separately by Schwinger \cite{Schwinger:1963} and Kibble \cite{Kibble:1963}, and was followed by a paper by Isham and Deser \cite{Deser:1976ay}, but their derivations are somewhat terse and incomplete. With modernised notation and understanding, the formalism has been developed further in many works across different communities-- too many to cite exhaustively here. Some details are often glossed over in the literature, and, while not necessarily affecting the final outcome, some intermediate steps lack detail or are sometimes incorrect. Some of the technical details were pointed out by Henneaux \cite{Henneaux:1978wlm,Henneaux:1983vi}, but appear to have been largely overlooked. We aim to fill some of these gaps by a detailed derivation of the vielbein form and identification of the symmetry generators. 

While often very convenient and geometrically well motivated, we will avoid the language of forms and frame bundles, primarily for pedagogical reasons; we will not provide a detailed account of the geometry or formal mathematical details on vielbeins, which is well covered in the literature. Similarly, the geometric foliation of spacetime into the $D{+}1$ formulation is well understood and covered pedagogically elsewhere, e.g. \cite{bojowald:2010, Gourgoulhon:2007ue}. Instead, we will take a straightforward field theory perspective, where we encode the metric degrees of freedom in a new field which happens to be a vielbein.

We will do so in different setups, convenient for different communities, hopefully bringing some clarity to some common sources of confusion. We aim to remain as general as possible but will, for simplicity and pedagogical reasons, ignore matter coupling and the cosmological constant. We omit the latter since its addition is trivial and the former to keep the formulae as short and simple as possible. With that said, we will try to keep many of the derivations explicit and retain intermediate steps for pedagogical reasons with a number of additional details in the appendix.

We will also remark here that, since we will be working with different formalisms, we will, in an attempt to simplify notation, reuse symbols in each of the sections instead of having different notations for the same objects with minor differences in formalism. We therefore advise \textbf{caution} when mixing equations from the different sections.

We will work in $d=D+1$ dimensions, in units where $c=\hbar=1$, and adopt the sign convention of Wald \cite{wald1984general}. There will be four different sets of indices, all subject to the summation convention. Greek letters $(\alpha, \beta, \dots, \mu, \nu, \dots)$ denote spacetime coordinate indices and run from $0$ to $D$. Uppercase Latin letters $(A, B, C,\dots)$ correspond to Lorentz indices and run from $0$ to $D$. The spatial parts of Lorentz and coordinate indices run from $1$ to $D$ and are denoted by lowercase letters from the top $(a,b,c, \dots)$ and middle $(i,j,k,\dots)$ of the Latin alphabet, respectively. 

We will use the mostly plus convention for the signature so that the Minkowski metric reads $\eta_{AB}= \mathrm{diag}(-1,+1,\dots,+1)$, and denote its spatial part as $\eta_{ab}=\delta_{ab}=\mathrm{diag}(+1,\dots,+1)$. We also use the totally antisymmetric epsilon symbol $\epsilon$, with $\epsilon_{01\dots D}=+1$ and $\epsilon_{1\dots D}=\epsilon^{1\dots D}=+1$, and the (anti-)symmetrisation bracket with a normalisation factor,
\begin{align}
    X_{(\alpha \beta)} = \tfrac{1}{2}\big[X_{\alpha \beta}+ X_{\beta \alpha} \big], && Y_{[\alpha \beta]} = \tfrac{1}{2}\big[Y_{\alpha \beta}- Y_{\beta \alpha} \big].
\end{align}

\section{\texorpdfstring{$D{+}1$}{D+1} Decomposition of the Vielbein and Einstein--Hilbert Action}

In this section, we start from the basic properties of a vielbein and a $D+1$ foliation of spacetime, and derive the corresponding vielbein decomposition in coordinates adapted to the foliation. The starting point is inspired by \cite{Peldan:1993hi} and produces the local Lorentz-covariant form of the $D+1$ decomposed vielbein, which we will later use to put the Einstein–Hilbert action into a canonical vielbein form.

A defining feature of a vielbein $e^A_{~\mu}(x)$ comes from its relation to the metric field given by
\begin{align}
\label{vielbein_def}
    g_{\mu \nu}(x)=e^A_{~ \mu}(x)\eta^{~}_{AB}e^B_{~ \nu}(x).   
\end{align}
We can also define the vielbein of the inverse metric,
\begin{align}
    g^{\mu \nu}(x) = e^{\mu}_{~ A}(x)\eta^{AB}e^\nu_{~ B}(x),
\end{align}
where the position of the indices, similar to the metric, indicates whether we are referring to the vielbein $e^A_{~ \mu}$ or its inverse $(e^{-1})^\mu_{~A}$ without ambiguity, since $(e^{-1})^\mu_{~A}=e^\mu_{~ A}= \eta_{AB}g^{\mu \nu}e^B_{~ \nu}$. The inverse obeys,
\begin{align}
\label{inv_e}
    e^A_{~ \mu}e^\mu_{~  B}=\delta^A_{B}, && e^\mu_{~  A}e^A_{~ \nu}=\delta^\mu_\nu .
\end{align}
In contrast to the metric, which has $d(d+1)/2$ independent components, the vielbein components $e^A_{~ \mu}$ correspond to a generic $d\times d$ invertible matrix, with $d^2$ independent parameters. The additional $d(d-1)/2$ components of the vielbein correspond to the choice of orthonormal frame in the internal Lorentz space $(\mathbb{R}^{1,D},\eta)$ which is spanned by the $d$ Lorentz vectors $ e^A_{~0},e^A_{~1},\dots,e^A_{~D}$. This means that the metric is invariant under a local Lorentz transformation $e^A_{~ \mu}(x)\mapsto \Lambda^A_{~ B}(x)e^B_{~ \mu}(x)$, where $\Lambda^A_{~ C}\eta_{AB}\Lambda^B_{~ D}=\eta_{CD}$, so that $\Lambda^A_{~ B}e^B_{~ \mu}$ and $e^A_{~ \mu}$ generate the same metric,
\begin{align}
    \big(e^C_{~ \mu}\Lambda^A_{~ C}\big)\eta_{AB}\big(\Lambda^B_{~ D}e^D_{~ \nu}\big)=e^C_{~ \mu}\eta_{CD}e^D_{~ \nu}= g_{\mu \nu}.
\end{align}

\subsection{\textit{D}+1 Vielbein}
\label{sec:D+1_vielbeins}
If we now foliate spacetime into a set of spatial hypersurfaces with a future-directed unit normal vector $n^\mu$ (see \cite{bojowald:2010} or \cite{Gourgoulhon:2007ue} for modern pedagogical derivations), the metric can conveniently be decomposed. If we also choose the coordinates adapted to the foliation, we can write the metric components in terms of a Riemannian spatial metric $\gamma_{ij}(x)$ on the $D$-dimensional hypersurfaces, a lapse function $N(x)$, and a spatial shift vector $N^i(x)$, so that the normal vector depends only on $N$ and $N^i$,
\begin{align}
\label{n_def}
    n^\mu = (1/N,-N^i/N), && n_\mu = (-N,0),  
\end{align}
so that $n_\mu \mathrm{d}x^\mu=-N\mathrm{d}t$ and the metric reads,
\begin{align}
    \mathrm{d}s^2 =g_{\mu \nu}\mathrm{d}x^\mu\mathrm{d}x^\nu = -N^2 \mathrm{d}t^2 +\gamma_{ij}\big(\mathrm{d}x^i+ N^i \mathrm{d}t\big)\big(\mathrm{d}x^j+ N^j \mathrm{d}t\big),
\end{align}
or, in block matrix notation,
\begin{align}
\label{3+1_metric}
    g_{\mu \nu}=\begin{pmatrix}
        -N^2+N_kN^k & N_j \\
        N_i & \gamma_{ij} \\
    \end{pmatrix}, && N_i = \gamma_{ij}N^j.
\end{align}
The inverse metric $g^{\mu \nu}$ can be similarly decomposed,
\begin{align}
    g^{\mu \nu} = \frac{1}{N^2}
    \begin{pmatrix}
    -1 & N^j \\
    N^i & N^2 \gamma^{ij}-N^i N^j
    \end{pmatrix},
\end{align}
where $\gamma^{ij}$ is the inverse of the spatial $D$-metric $\gamma_{ij}$, $\gamma^{ik}\gamma_{kj}=\delta^i_j$, but in contrast to $\gamma_{ij}=g_{ij}$, it is not equal to $g^{ij}$. 

With the metric decomposed as \eqref{3+1_metric}, we can express the vielbein in terms of the lapse, shift, and spatial metric through \eqref{vielbein_def}. By direct substitution, we obtain the relations,
\begin{subequations}
\begin{align}
\label{timetime}
    e^{A}_{~ 0}\eta_{AB}e^B_{~ 0}&=-N^2 + N^k\gamma_{kl}N^l, \\
\label{vectorial}
    e^{A}_{~ i}\eta_{AB}e^B_{~ 0}&= \gamma_{ik}N^k, \\ 
\label{spatial}
    e^{A}_{~ i}\eta_{AB}e^B_{~ j}&= \gamma_{ij}.
\end{align}
\end{subequations}
For later convenience, we will adopt the notation $E^A_{~ i}=e^A_{~ i}$, so that \eqref{spatial} yields,
\begin{align}
\label{spatial_vielbein}
    E^A_{~ i}\eta_{AB}E^B_{~ j}= \gamma_{ij}.
\end{align}
Since the foliation is chosen so that $\gamma_{ij}$ is Riemannian and non-degenerate (i.e. $\det(\gamma)\neq 0$), $\{E^A_{~ i}\}$ forms a set of $D$ linearly independent spacelike Lorentz vectors spanning the spatial hypersurfaces. However, from \eqref{timetime}, it is not necessary for $e^A_{~ 0}$ to be timelike, as $g_{00}$ can have any sign, nor that it is orthogonal to $E^A_{~ i }$ since generically $E^A_{~i}\eta_{AB}e^B_{~0}=N_i\neq0$.\footnote{Note that $d$ spacelike vectors can span the whole Lorentz space. For example, in $1{+}1$ dimensions with $\eta=\mathrm{diag}(-,+)$, let $\hat e_0=(1,0),\;\hat e_1=(0,1)$. Define $e_0=\hat e_0+\lambda \hat e_1, \; e_1=\hat e_0-\lambda \hat e_1,\; \lambda>1$. Then $\eta(e_0,e_0)=\eta(e_1,e_1)=-1+\lambda^2>0$, so $e_0,e_1$ are spacelike, yet linearly independent and span the whole Lorentz space. Their span nevertheless contains timelike vectors since $e_0+e_1=2\hat e_0$ is timelike.} Although we could in principle work with $e^A_{~ 0}$, it is often more convenient to introduce a normalised Lorentz vector $X^A$ orthogonal to $E^A_{~i}$, which necessarily, in contrast to $e^A_{~ 0}$, is timelike and satisfies,
\begin{align}
\label{X_def}
    X_A E^A_{~ i}=0, && X_A X^A=-1,
\end{align}
where we have used $\eta_{AB}$ to lower the Lorentz indices, $X_A=\eta_{AB}X^B$. The orthogonality condition makes $X^A$ unique up to a sign, which determines whether $X^A$ is future- or past-directed and will be determined below.

The vectors $X^A$ and $E^A_{~ i} $ are linearly independent and span the Lorentz space, and thus constitute a basis, implying that $e^A_{~ 0}$ can be written as a linear combination,
\begin{align}
    e^A_{~\; 0}= a X^A +  E^A_{~ i}\,b^i,
\end{align}
for some Lorentz scalars $a(x)$ and $b^i(x)$. We can now relate these to the $D+1$-variables by substituting the above into \eqref{vectorial},
\begin{align}
    E_{A i}\big(a X^A+b^k E^A_{~ k}\big)&
    =\gamma_{ik}b^k=\gamma_{ik}N^k \qquad \Longrightarrow \qquad b^j=N^j,
\end{align}
where we have used $X_A E^A_{~\, i}=0$ and $E_{A i}E^A_{~ k}=\gamma_{ik}$. Using this  together with \eqref{timetime} yields the relation:
\begin{align}
    \big(aX_A+ N^iE_{Ai}\big)\big(a X^A+ N^jE^A_{~ j}\big)=-a^2 + N^i\gamma_{ij}N^j=-N^2+ N^i \gamma_{ij}N^j,
\end{align}
determining $a^2=N^2$, leaving the ambiguity $a=\pm N$, reflecting the freedom to choose $X^A$ as future- or past-directed. Requiring $X^A$ to be future-directed relative to the chosen time coordinate amounts to imposing $X_A e^A_{~0}=-a<0$, which fixes $a=N>0$. We can now express $e^A_{~ \mu}$ in terms of $N, N^i, X^A$ and $E^A_{~ i}$,
\begin{align}
\label{e_decomp}
    e^A_{~\; 0} &= N X^A + N^i E^A_{~ i}, && e^A_{~ i}= E^A_{~ i}.
\end{align}
We can relate the internal time direction given by $X^A$ to the unit vector normal to the spatial hypersurface, $n^\mu$, by noting that,
\begin{align}
\label{X_n_relation}
    e^A_{~ \mu}n^\mu = e^A_{~ 0}n^0 +E^A_{~ i} n^i = \big(NX^A + N^i E^A_{~ i}\big)\tfrac{1}{N}- E^A_{~ i}\tfrac{N^i}{N} = X^A,
\end{align}
which immediately yields the relations,
\begin{align}
    X^A\eta_{AB}X^B= n^\mu n^\nu g_{\mu \nu}=-1, && X_A E^A_{~ i} = n^\mu g_{\mu i}=0.
\end{align}
We could equally well have used \eqref{X_n_relation} to impose that the spacetime and internal temporal directions are aligned, thereby determining the sign of $a$.

We can also define the inverse spatial vielbein  $E^j_{~ A}= \eta_{AB}\gamma^{ji}E^B_{~ i}$, obeying,
\begin{align}
    E^A_{~\, i}E^j_{~ A}=\delta^i_j.
\end{align}
Note that $E^i_{~ A}$ is not the $(i,A)$ components of the inverse $e^{\mu}_{~ A}$, so while the above is the inverse in Lorentz space, if we instead contract the coordinate indices, we do not get $\delta^A_B$, but rather the projector $P^A_B$ onto the subspace orthogonal to $X^A$, i.e. the spatial Lorentz space,
\begin{align}
\label{proj}
    E^A_{~i}E^i_{~ B}= \delta^A_B + X^AX_B=P^A_B, && P^A_BP^B_C=P^A_C, && P^A_B X^B =0.
\end{align}
Using $E^j_{~ A}$ and \eqref{inv_e} we can decompose the inverse vielbein $e^\mu_{~ A}$ as,
\begin{align}
\label{einv_decomp}
    e^0_{~ A} &= -N^{-1}X_A, \quad\quad &&e^i_{~ A}= E^i_{~ A}+ N^{-1}N^iX_A,
\end{align}
where in particular, $e^i_{~ A}\neq E^i_{~ A}$.

We will later make explicit choices for the parametrisation of $X^A$, but it is important to note that it is independent of the lapse and shifts, and does not provide any extra degrees of freedom beyond those of the vielbein. This can be made explicit by the fully Lorentz-covariant parametrisation that depends only on $E^A_{~ i}$ \cite{Castellani:1981ue},
\begin{align}
\label{X_hodge}
    X_A &= -\frac{1}{D!\sqrt{\gamma}}\epsilon_{AB_1\dots B_D}\epsilon^{i_1\dots i_D}E^{B_1}_{~ i_1}\dots E^{B_D}_{~ i_D}.
\end{align}
From which it is clear that $X_A$ does not provide any additional degrees of freedom apart from the variables $N, N^i, E^A_{~ i}$, corresponding to precisely $d^2$ independent fields of a vielbein. It also implies that a choice of $X_A$ fixes components of $E^A_{~ i}$ or vice versa, e.g. the gauge choice $X^A= (1,0)$ fixes the so-called time gauge $E^0_{~ i}=0$.

With the general decomposition of the vielbein \eqref{e_decomp} established, we are now ready to decompose the Einstein–Hilbert action and transition to the canonical vielbein formulation. However, there is no unique way to do this. For example, different choices of $X^A$, boundary terms, or canonical variables yield different canonical structures. We will cover two different approaches, starting with a fully Lorentz-covariant formulation and later the SO($D$)-covariant form.

\subsection{\textit{D}+1 Decomposition of the Einstein–Hilbert action}
\label{sec:D+1_EH}
Deriving the $D+1$ form of the Einstein–Hilbert action is done in many places in the literature and can be done in different ways. We will therefore only cover the basics and point out the aspects that will be important to us when we later transition to the canonical vielbein formulation.

Let us start with the Einstein–Hilbert action in $d$ dimensions,
\begin{align}
\label{EH}
    \Scal= \mpl \dint x\,\sqrt{|g|}\,\rd \!R, && g = \det g_{\mu \nu}.
\end{align}
To decompose this into its $D+1$ form, we use the Gauss-Codazzi equation, which relates the $d$-Ricci scalar $\rd \! R$ to the $D$-Ricci scalar $\rD \! R$ and the extrinsic curvature $K_{ij}$, with an additional divergence term,
\begin{align}
\label{Gauss-Codazzi}
    \rd \!R= \rD\! R + K^{ij}K_{ij}-K^2-2\,\rd\nabla_\mu v^\mu,   && v^\mu = n^\nu\, \rd\nabla_\nu n^\mu - n^\mu \,\rd\nabla_\nu n^\nu.
\end{align}
$\rd\nabla_\mu$ is the $g_{\mu \nu}$-compatible Levi-Civita connection $(\rd\nabla_\mu g_{\alpha \beta}=0)$, and $n^\mu$ is the timelike unit normal vector \eqref{n_def}. The extrinsic curvature $K_{ij}$ is defined by the Lie derivative of the spatial metric along $n^\mu$,
\begin{align}
    K_{ij}=\tfrac{1}{2}\Lcal_{\vec{n}} \gamma_{ij}= \frac{1}{2N}\Big[\dot{\gamma}_{ij}- 2\,\rD\nabla^{}_{(i}N_{j)}^{\,} \Big], && K = \gamma^{ij}K_{ij}.
\end{align}
$\rD\nabla_i$ is the $\gamma_{ij}$-compatible Levi-Civita connection ($\rD\nabla_i \gamma_{jk}=0$) which is used to construct the spatial curvature $\rD \! R$. The expressions for the $d$-dimensional Christoffel symbols $\rd\!\Gamma^\mu_{~ \alpha \beta}$ of $\rd\nabla$ and the relation to the spatial Christoffel symbols $\rD\Gamma^i_{~ jk}$ of $\rD \nabla_i$ can be found in Appendix \ref{sec:Christoffel}.

When \eqref{Gauss-Codazzi} is substituted into the action \eqref{EH}, $\sqrt{|g|}\rd\nabla_\mu v^\mu = \partial_\mu(\sqrt{|g|}v^\mu)$ produces a boundary term. While boundary terms do not affect the field equations in the bulk, they do contribute to the canonical momenta, and therefore change the canonical structure.\footnote{The boundary terms are also part of specifying a well-posed variational principle, and enters the definition of gravitational surface charges at infinity, e.g. ADM energy and the Bondi mass. So different choices shift the explicit charge expressions and their values without changing the bulk equations.} For example, since the vector $v^\mu$ depends on derivatives of $n^\mu$ and thus $\dot{N}$ and $\dot{N}^i$, the term would affect their canonical momenta $P=\delta \Scal/\delta \dot{N}$, $P_i=\delta \Scal/\delta \dot{N}^i$. We will, therefore, add the “temporal York–Gibbons–Hawking” counterterm so that we can neglect $\rd\nabla_\mu v^\mu$. This will result in an particularly useful form of the action which is independent of $\dot{N}$ and $\dot{N}^i$,
\begin{align}
\label{EH_decomp}
    \Scal= \mpl \dint x\, N\sqrt{\gamma}\Big [\!\rD \! R+ K^{ij}K_{ij}-K^2\Big]. 
\end{align}
Here $\gamma= \det \gamma_{ij}$, $\det g_{\mu \nu} = -N^2\det \gamma_{ij} $ and we emphasise that the spatial curvature $\rD \! R$ is constructed from $\rD\nabla_i$ which is a function purely of the spatial metric $\gamma_{ij}$, and its spatial derivatives. In contrast, the extrinsic curvature contains $\dot{\gamma}_{ij}$, but no time derivatives of the lapses and shift. This is one of the reasons why the decomposition \eqref{3+1_metric} is useful, as it restricts the dynamical content to the spatial metric. 

While the regular metric $D+1$ formalism now introduces the canonical momenta conjugate to the spatial metric $\gamma_{ij}$, we will instead substitute for the vielbeins and consider the fields $N,N^i, E^A_{~ i}$ as canonical variables. By direct substitution of \eqref{spatial_vielbein} into the spatial metric,
\begin{align}
\label{gamma_dot}
    \dot{\gamma}_{ij}= 2\eta_{AB}E^A_{~(i}\dot{E}^B_{~ j)}\,,
\end{align}
the extrinsic curvature takes the form,
\begin{align}
\label{K_vielbein}
    K_{ij}= \frac{1}{N}\Big[E^A_{~(i}\eta_{AB}\dot{E}^B_{~ j)}-\rD \nabla_{(i} N_{j)}\Big].
\end{align}
We could in principle also write the spatial curvature $\rD \! R$ explicitly in terms of the spatial vielbeins $E^A_{~ i}$, but since it will only depend on spatial derivatives,
\begin{align}
    \partial_i \gamma_{jk}= 2\eta_{AB}\partial_i E^A_{~(j}E^B_{~k)},
\end{align}
which will not affect the canonical structure, we keep it implicit, but emphasise that it, as a function of only $\gamma_{ij}$, is locally Lorentz invariant.

With the action in the form \eqref{EH_decomp}, we are ready to transition to the canonical formulation, but as stressed before, there is no unique way of doing this. In the next section, we will work in a manifestly Lorentz-covariant formulation and use $E^A_{~ i}$ as canonical fields.

\section{Lorentz-covariant Canonical Formulation}
\label{sec:Lorentz_covariant}

In this section, we derive the manifestly Lorentz-covariant form of the vielbein phase-space action, $\Scal= \dint x\big[\pi^i_{A}\dot{E}^A_{~i} - \mathcal{H}(E, \pi)\big]$, from which the transition to the Hamiltonian formulation is straightforward, and we can derive the symmetry generators and constraints of the theory.

\subsection{Lorentz invariant phase-space action}
\label{sec:Phase_space_Action}

The first step is to define the momenta conjugate to $E^A_{~ i}$,
\begin{align}
\label{pi}
    \pi^i_A &= \frac{\delta \Scal}{\delta \dot{E}^A_{~ i}} = 2\mpl \sqrt{\gamma}E_{Aj}\Big[ K^{ij}-\gamma^{ij}K\Big].
\end{align}
Note that not all $d(d-1)$ components of $\pi^i_A$ are independent, but the combination $\Jcal^{AB}=\pi_{\phantom{i}}^{i[A}E{}^{B]}_{~ i}$ vanishes due to the fact that $K^{ij}$ \eqref{K_vielbein}, and $\gamma^{ij}$ are symmetric. From the condition $X_AE^A_{~j}=0$ \eqref{X_def}, it also follows that $X^A\pi^i_{A}=0$, so that $\pi^i_A$ is a spatial Lorentz covector. However, since,
\begin{align}
    2X_{A}E^i_{~ B}\Jcal^{AB}=\pi^i_AX^A=0,
\end{align}
follows from $\Jcal^{AB}=0$ and $X_AE^A_{~i}=0$, the orthogonality $\pi^i_A X^A=0$ is not an independent constraint. These equations follow from the definition of the momenta and therefore yield primary constraints:
\begin{align}
\label{JAB}
    \Jcal^{AB}=\pi^{i[A}_{\phantom{i}}E^{B]}_{~ i}=0.
\end{align}
These $d(d{-}1)/2$ conditions constrain the $d(d{-}1)$ components of $\pi^i_{A}$ to only $d(d{-}1)/2$ independent fields, in agreement with the degrees of freedom of $\dot{\gamma}_{ij}$. The constraints arise due to the local Lorentz invariance of the action \eqref{EH_decomp}, or more explicitly, the fact that the action is independent of the $d(d{-}1)/2$ Lorentz degrees of freedom contained in $E^A_{~ i}$. In Appendix \ref{sec:Lorentz_momenta} we show explicitly how $\Jcal^{AB}$ corresponds to the canonical momenta conjugate to the Lorentz components of $E^A_{~ i}$, and $\Jcal^{AB}=0$ thus constitutes the primary constraints corresponding to the fact that the Lorentz fields are non-dynamical.

Although we, in principle, could use the notion of a pseudoinverse to invert \eqref{pi} for the velocities $\dot{E}^A_{~i}$, we instead proceed with a different approach and see that the explicit expression for the velocity is not needed. For completeness, we present the inversion in terms of the Moore–Penrose pseudoinverse in Appendix \ref{pseudo_inv_method}.

Working off the primary constraint surface and with an unconstrained $\pi^i_A$, we can invert \eqref{pi} for the extrinsic curvature,
\begin{align}
\label{Kij}
    \Kcal^{i}_{~j}&=\frac{1}{2\mpl\sqrt{\gamma}}\Big[\pi^{i}_{A}\,E^{A}_{~j}-\tfrac{1}{d-2}\delta^{i}_{j}\,\pi^{k}_{A}\,E^{A}_{~k}\Big].
\end{align}
Here we have used $\Kcal_{ij}(E, \pi)$ to distinguish it from the manifestly symmetric $K_{ij}(E, \dot{E})$, as the inverted expression contains a skew-symmetric part:
\begin{align}
\label{K_antisym}
    \Kcal_{[ij]}=\frac{-1}{2\mpl\sqrt{\gamma}}E_{A i}E_{B j}\Jcal^{AB},
\end{align}
which vanishes on the primary constraint surface \eqref{JAB}. We can now use $K_{ij}(E, \pi) = \Kcal_{(ij)}$ for the symmetric expression that appears in the Einstein–Hilbert action \eqref{EH_decomp}.

We will now use a trick to derive a Lorentz-covariant expression for the canonical one form $\pi^i_A\dot{E}^A_{~ i}$, by considering the $e^A_{~\mu }$-compatible covariant derivative $\rd\Dcal_\mu$ of $e^A_{~\mu}$ to evaluate the covariant time derivative of $E^A_{~i}$. By definition,
\begin{align}
    \rd\Dcal_\mu e^A_{~ \nu}= \partial_\mu e^A_{~ \nu} - \rd\!\Gamma^\sigma_{~ \mu \nu}e^A_{~ \sigma}+ \rd\!\omega_{\mu}{}^{A}_{~B}\, e^B_{~ \nu}=0,
\end{align}
where $\rd \!\omega_{\mu A B}=e^\nu_{~[A} \rd\nabla_{\mu}e^{}_{B] \nu}$ is the spin connection and $\rd\!\Gamma^\mu_{~ \alpha \beta}$ are the standard metric Christoffel symbols. 

Consider now the components $(\mu, \nu)=(0,i)$, which yield the covariant time derivative of the spatial vielbein, which with the $D+1$-decomposition of the vielbein \eqref{e_decomp} and \eqref{einv_decomp}, can be written,
\begin{align}
    \rd\Dcal_0 e^A_{~ i} = \dot{E}^A_{~ i}- \big[\rd\!\Gamma^0_{~ 0i}N^j + \rd\!\Gamma^j_{~ 0 i}\big]\,E^A_{~ j} - N\rd\!\Gamma^0_{~ 0 i}X^A+ \rd\!\omega_{0}{}^{A}_{~B}E^B_{~ i}=0.
\end{align}
Using the standard $D+1$ expressions for the Christoffel symbols, which are available in Appendix \ref{sec:Christoffel}, the expression reads,
\begin{align}
\label{E_dot}
    \dot{E}^A_{~ i}- \big[NK^j_{~ i}+ \rD\nabla_i N^j\big]\,E^A_{~ j}-\big[\partial_i N + K_{ij}N^j\big]\,X^A+ \rD\!\omega_{0}{}^{A}_{~B} E^B_{~ i}=0.
\end{align}
Contracting with the unconstrained $\pi^i_A$ and substituting the spin connection $\rd\!\omega_{0}{}^{A}_{~B}$ from \eqref{w0AB}, yields\footnote{Note that we could equivalently take the time derivative of $X_A E^A_i=0$ to solve for $\dot{X}_A$ in terms of $X_A$ and $\dot{E}^A_{~ i}$, thereby eliminating the time derivative of $X_A$.}
\begin{align}
\label{can_form}
    \pi^i_A \dot{E}^A_{~ i} - N \pi^i_AE^A_{~ j}K^j_{~ i}-\pi^i_AE^A_{~j}\rD\nabla_i N^j +\Big[E^i_{[A}\dot{E}^{}_{B] i}-X_{[A}\dot{X}_{B]}- E^i_{[A}E^j_{B]}\rD\nabla_i N_j \Big]\Jcal^{AB}=0.
\end{align}
Since the expressions \eqref{E_dot} and $\pi^i_A$ transform covariantly under local Lorentz rotations, this expression is a Lorentz scalar. However, we emphasise that the canonical one-form $\pi^i_A \dot{E}^A_{~ i}$ is not; rather, its non-homogeneous transformation is cancelled by $\rD\!\omega_{0}{}^{A}_{~B}E^B_{~ i}\pi^i_{A}$.

Using \eqref{Kij} we can express the Einstein–Hilbert Lagrangian \eqref{EH_decomp} in terms of $E^A_{~i}, \pi^i_A, N$ and $N^i$, and isolate the term $NK^j_{~ i}\pi^i_{A}E^A_{~j}$,
\begin{align}
    \Lcal=N\mpl \sqrt{\gamma}\Big[\!\rD \!R+K_{ij}K^{ij}-K^2 \Big] =NK^j_{~ i}\pi^i_{A}E^A_{~j}+ N\mpl \sqrt{\gamma}\Big[\rD \! R-K_{ij}K^{ij}+K^2 \Big],
\end{align}
which also appears in \eqref{can_form}. Substituting this term from \eqref{can_form} yields the Lagrangian in terms of the canonical one-form,
\begin{align}
    \Lcal&= \pi^i_A \dot{E}^A_{~ i}+N\mpl \sqrt{\gamma}\Big[\!\rD \! R-\Kcal_{ij}\Kcal^{ij}+\Kcal^2 \Big]-\pi^i_{A}E^A_{~j}\rD\nabla_i N^j + W_{AB}\Jcal^{AB},
\end{align}
where we have substituted $K_{ij} = \Kcal_{(ij)}$ for $\Kcal_{ij}- \Kcal_{[ij]}$, thereby generating an additional term proportional to $\Jcal^{AB}$ \eqref{K_antisym} and collected the terms proportional to $\Jcal^{AB}$ into the variable $W_{AB}$,
\begin{align}
\label{Omega_def}
    W_{AB}&=E^i_{[A}\dot{E}^{}_{B] i}-X_{[A}\dot{X}_{B]}- E^i_{[A}E^j_{B]}\rD\nabla_i N_j -\tfrac{N}{2}E^i_{[A}E^j_{B]}\Kcal_{ij}.
\end{align}
Substituting this into the action and partially integrating the term $\pi^i_{A}E^A_{~j}\rD\nabla_i N^j$, we get,
\begin{align}
\label{can_action}
    \Scal =\dint x \Big[ \pi^i_A\dot{E}^A_{~ i} + N\Rcal+ N^i \widetilde{\Rcal}^i +W_{AB}\Jcal^{AB}\Big],
\end{align}
where we have defined,
\begin{subequations}
\begin{align}
\label{vielbein_ham}
    \Rcal &= \mpl \sqrt{\gamma}\,\rD \!R+ \tfrac{1}{4\mpl \sqrt{\gamma}}\Big[\tfrac{1}{d-2}\big(\pi^i_AE^A_{~ i}\big)^2-E^A_{~ j}\pi^j_{~ B}E^{B}_{~ i}\pi^i_{ A}\Big],\\[2mm]
\label{old_mom_const}
    \widetilde{\Rcal}_i&=E^A_{~ i}\rD\Dcal_j \pi^j_{~ A}=E^A_{~i}\Big[\partial_j \pi^j_{ A}-\rD\!\omega_{jA}{}^B \pi^j_{B} \Big],
\end{align}
\end{subequations}
where $\rD\Dcal_i$ is the $E^A_{~ i}$-compatible spatial covariant derivative, $\rD\Dcal_i E^A_{~ j}=0$, with corresponding Christoffel and spin connection, $\rD\Gamma^i_{~ j k}$ and $\rD\!\omega_{iAB}$ (form available in Appendix \ref{sec:Christoffel}). Note that $\pi^i_A$ is a spacetime vector density of weight 1, so that its covariant derivative must be computed as:
\begin{align}
    \rD \nabla_i \pi^j_A = \partial_i \pi^j_A + \rD\!\Gamma^j_{~ik}\pi^k_{A}-\rD\!\Gamma^k_{~ik}\pi^j_A.
\end{align}
The action \eqref{can_action} a priori looks different from the metric formulation, even if the relations for the spatial metric and its momenta, $\gamma_{ij}= E_{Ai}E^A_{~j}$ and $\pi^{ij}= \tfrac{1}{2}\pi^{(i}_{A}E^{j)A}$, are used. In Appendix \ref{sec:metric_action} we compare the different terms and show that this form is strictly equal to the common metric phase-space action $\dint x [\pi^{ij}\dot{\gamma}_{ij}+N\Rcal^\gamma+N^i \Rcal_i^\gamma]$, but also that the different forms of  $\Rcal$ and $\widetilde{\Rcal}_i$ are relevant for the transformations they generate on the vielbein variables.

\subsection{Hamiltonian Formulation}
With the action in phase-space form \eqref{can_action}, the transition to the Hamiltonian formulation is straightforward. Since we want to treat all variables on equal footing, we also introduce canonical momenta conjugate to the lapse and shift,
\begin{align}
    P = \frac{\delta \Scal}{\delta \dot{N}}, && 
    P_i = \frac{\delta \Scal}{\delta \dot{N}^i}.
\end{align}
Because we have disregarded the divergence term in \eqref{Gauss-Codazzi}, the action is independent of $\dot{N}$ and $\dot{N}^i$, so these momenta vanish identically. Together with $\Jcal^{AB}=0$, we therefore obtain the primary constraints,
\begin{align}
\label{primary}
    P=0, && P_i=0, && \Jcal^{AB}=0.
\end{align}
With the canonical pairs $(N, P),\, (N^i, P_j), \,(E^A_{~ i}, \pi^j_B),$ we can define the canonical Poisson brackets,
\begin{subequations}
\label{poisson}
\begin{align}
    \{N(x),P(y)\}&=\delta(x-y),\\
    \{N^i(x),P_j(y)\}&=\delta^i_j\delta(x-y),\\
    \{E^A_{~ i}(x),\pi^j_{ B}(y)\}&=\delta^A_B \delta^j_i\delta(x-y),
\end{align}
\end{subequations}
where all other brackets between the canonical variables vanish and $\delta(x)$ is the $D$-dimensional Dirac delta.

The total Hamiltonian is now simply obtained by adding the primary constraints \eqref{primary} with Lagrange multipliers to the Legendre transform of the Einstein–Hilbert Lagrangian \eqref{can_action},
\begin{align}
\label{tot_ham}
    H_T = -\Dint x \Big[
        N \Rcal+ N^i \widetilde{\Rcal}_i 
        + \lambda P + \lambda_i P^i + \widetilde{\lambda}_{AB}\Jcal^{AB} 
    \Big].
\end{align}
Since the Lagrange multiplier $\widetilde{\lambda}_{AB}$ is arbitrary, we have absorbed $W_{AB}$ into it without loss of generality, leaving the Hamiltonian independent of velocities $\dot{E}^A_{~ i}$.\footnote{In principle $\widetilde{\lambda}_{AB}$ should not be functions of the velocities, but this can be remedied by using the form \eqref{Edot_sol}, although this will not be needed in the constraint analysis.}

The above expression is perfectly acceptable, but it is not yet in the most useful form. In particular, $\widetilde{\Rcal}_i$ does not generate pure spatial diffeomorphisms on $(E^A_{~i},\pi^i_A)$. In Appendix \ref{sec:mod_gen_1}, the proper generator is constructed and shown to differs from $\widetilde{\Rcal}_i$ only by a term proportional to $\Jcal^{AB}$. Explicitly,
\begin{align}
\label{new_momentum}
    \Rcal_i &= \widetilde{\Rcal}_i + \rD\!\omega_{iAB}\Jcal^{AB}.
\end{align}
The Hamiltonian can be written in terms of $\Rcal_i$, by a simple shift of the Lagrange multiplier,
\begin{align}
    \lambda_{AB}=\widetilde{\lambda}_{AB}-N^i\,\rD\!\omega_{iAB},
\end{align}
yielding the final expression,
\begin{align}
\label{Htot}
    H_T = -\Dint x \Big[
        N \Rcal+ N^i \Rcal_i 
        + \lambda P + \lambda_i P^i + \lambda_{AB}\Jcal^{AB} 
    \Big].
\end{align}
We can now derive the Hamiltonian field equations for the vielbein and its momentum in the standard way,
\begin{subequations}
\begin{align}
    \dot{E}^A_{~ i} &= \{E^A_{~ i}, H_T\}\approx NK^j_{~i} E^A_{~ j}+ \Lcal_{\vec{N}}E^A_{~ i}- \lambda^A_{~ B}E^B_{~ i}\\
    \dot{\pi}^i_{A} &= \{\pi^i_{A}, H_T\} \approx \frac{\delta \Rcal[N]}{\delta E^A_{~ i}} +\Lcal_{\vec{N}}\pi^i_{A} + \pi^i_B\lambda^B_{~ A},
\end{align}
\end{subequations}
where $\Rcal[N]=\Dint x\,N(x) \Rcal(x)$, and $\Lcal_{\vec{N}}$ is the Lie derivative in the $N^i$ direction (see \eqref{diff}) and,
\begin{align}
    \frac{\delta \Rcal[N]}{\delta E^A_{~i}}&=-2\mpl \sqrt\gamma\,E_{Aj}\Big[N\Big(\!\rD \!R^{ij}-\tfrac{1}{2}\gamma^{ij}\rD \!R\Big)+\gamma^{ij}\rD\nabla^2 N-\rD\nabla^i\rD\nabla^j N\Big]\notag\\
    &\quad\,-N\Big[\Kcal^i_{~j}\delta^B_A -\tfrac{1}{2}E^i_{~A}E^B_{~k}\Kcal^k_{~j} \Big]\pi^j_{~B}.
\end{align}

The equations of motion for the lapse and shifts instead yield,
\begin{align}
    \dot{N}= \{N, H_T\}=-\lambda, && \dot{N}^i= \{N^i, H_T\}=-\lambda^i,
\end{align}
which after gauge fixing, determines $\lambda$ and $\lambda^i$, while their momenta will provide constraints, which we will analyse in the next section. The antisymmetric contraction $E_{Ai}\dot{E}^A_{~j}$ can then be used to determine $\lambda_{AB} \approx \widetilde{\lambda}_{AB}+ W_{AB}- N^k \omega_{kAB}$ for the arbitrary $\widetilde{\lambda}_{AB}$ introduced in \eqref{tot_ham}.

\subsection{Constraints and Symmetry Operators}
\label{sec:sym_op_1}
With the Hamiltonian in the form \eqref{Htot}, the constraint analysis and derivation of the constraint algebra are straightforward. Following the Dirac--Bergmann algorithm, we require that the primary constraints are preserved in time. For the lapse and shift momenta $P$ and $P_i$ this directly yields,
\begin{subequations}
\begin{align}
\label{ham_const}
    \dot{P}(x) &= \{P(x),\, H_T\}\approx \Rcal(x) \,\approx0,\\
\label{mom_const}
    \dot{P}_i(x) &= \{P_i(x), H_T\}\approx \Rcal_i(x) \approx0,
\end{align}
\end{subequations}
which produces the vielbein version of Hamiltonian and momentum constraints. We also have the primary constraints $\Jcal^{AB}=0$, which must be preserved in time, i.e. $\dot{\Jcal}^{AB}\approx 0$, where,
\begin{align}
    \{\Jcal^{AB}(x), H_T\}\approx 
    -\Dint y \Big[
        N(y)\{\Jcal^{AB}(x),\Rcal(y)\}
        &+N^i(y)\{\Jcal^{AB}(x),\Rcal_i(y)\}\notag\\
\label{Jdot}
        &+\lambda_{CD}(y)\{\Jcal^{AB}(x),\Jcal^{CD}(y)\}
    \Big].
\end{align}
To evaluate these brackets, it is convenient to smear the operator $\Jcal[\theta]$ by the antisymmetric parameters $\theta_{AB}(x)=-\theta_{BA}(x)$,
\begin{align}
\label{Lorentz_gen}
    \Jcal[\theta]= \Dint x\; \theta_{AB}(x)\Jcal^{AB}(x),
\end{align}
which generates infinitesimal local Lorentz rotations $\Lambda^A_{~ B}= \delta^A_{ B}+ \theta^A_{~ B}+ \dots$ of the spatial vielbein and its momenta through the Poisson bracket,
\begin{align}
    \{E^A_{~ i}(x), \Jcal[\theta]\}= \theta^A_{~ B}(x)E^B_{~ i}(x), &&
    \{\pi^i_A(x), \Jcal[\theta]\}= -\pi^i_B(x)\theta^B_{~ A}(x).
\end{align}
With this, a straightforward computation yields,
\begin{subequations}
\label{J_brackets}
\begin{align}
\label{so3+1_lie_algebra}
    \{\Jcal^{AB}(x),\,\Jcal^{CD}(y)\}
    &=\Big[\eta^{A[C}\,\Jcal^{D]B}(x)-\eta^{B[C}\,\Jcal^{D]A}(x)\Big]\,\delta(x-y),\\
\label{JR_i}
    \{\Jcal^{AB}(x),\Rcal_i(y)\}
    &= \Jcal^{AB}(y)\, \pdv{}{y^i}\delta(x-y),\\
\label{JR}
    \{\Jcal^{AB}(x), \Rcal(y)\}
    &=0,
\end{align}
\end{subequations}
where the first line is the Lorentz algebra, \eqref{JR_i} corresponds to a spatial diffeomorphism of the local Lorentz generator, and \eqref{JR} shows that the Hamiltonian constraint is Lorentz invariant. Substituting \eqref{J_brackets} into \eqref{Jdot}, $\dot{\Jcal}^{AB}$ vanish weakly on the primary constraint surface, and does not generate any secondary constraints.

According to Dirac and Bergmann, we must also impose that the secondary constraints $\Rcal \approx0$ and $\Rcal_i\approx0$ remain constant in time,
\begin{subequations}
\begin{align}
\label{R_dot}
    \!\!\!\dot{\Rcal}(x) &\approx \!-\Dint y\Big[N(y) \{\Rcal(x)\;,\Rcal(y)\}+N^j(y) \{\Rcal(x)\;,\Rcal_j(y)\} + \lambda_{AB}(y) \{\Rcal(x)\;,\Jcal^{AB}(y)\} \Big]\approx0,\\
\label{Ri_dot}
    \!\!\!\dot{\Rcal}_i(x) &\approx\! -\Dint y\Big[N(y) \{\Rcal_i(x),\Rcal(y)\}+N^j(y) \{\Rcal_i(x),\Rcal_j(y)\} + \lambda_{AB}(y) \{\Rcal_i(x),\Jcal^{AB}(y)\} \Big]\approx0.
\end{align}
\end{subequations}
From this, using \eqref{JR_i} and \eqref{JR}, we can conclude that the last brackets in each row vanish weakly. By explicit computations--in direct analogy with the ones in the metric formulation--(see \cite{Khoury:2011ay} Appendix A), the remaining brackets can be computed,
\begin{subequations}
\label{diff_alg}
\begin{align}
\label{RR}
     \{\Rcal(x),\Rcal(y)\}
     &= \left[\Rcal^i(y)\pdv{}{y^i}-\Rcal^i(x)\pdv{}{x^i}\right]\delta(x-y),\\
\label{RR_i}
     \{\Rcal(x),\Rcal_i(y)\}
     &=\Rcal(y)\pdv{}{y^i}\delta(x-y),\\
\label{R_iR_j}
    \{\Rcal_i(x), \Rcal_j(y)\}
    &=\left[\Rcal_i(x) \pdv{}{y^j}-\Rcal_j(y)\pdv{}{x^i}\right]\delta(x-y),
\end{align}
\end{subequations}
where $\Rcal^i = \gamma^{ij}\Rcal_j$. The above brackets are all proportional to $\Rcal$ and $\Rcal_i$, and are thus weakly zero. This implyies that \eqref{R_dot} and \eqref{Ri_dot} both vanish weakly without implying any further constraints, thereby halting the Dirac–Bergmann algorithm.

The brackets \eqref{J_brackets} and \eqref{diff_alg} form a first-class algebra which, written in terms of the smeared operators \eqref{Lorentz_gen} and,
\begin{align}
    \Rcal[\chi] = \Dint x\, \chi(x) \Rcal(x), && \vec{\Rcal}[\vec{\xi}] = \Dint x\, \xi^i(x) \Rcal_i(x),
\end{align}
takes the form,
\begin{subequations}
\label{full_alg}
\begin{align}
    \{ \Jcal[\theta],\Jcal[\theta']\}&=\Jcal\big[[\theta,\theta']\big], 
    & [\theta,\theta']_{AB}&=\theta_{AC}\theta'^{C}_{~B}-\theta'_{AC}\theta^C_{~B},\\
    \{\Jcal[\theta],\vec{\Rcal}[\vec{\xi}]\}&=\Jcal[\Lcal_{\vec{\xi}}\theta], & \Lcal_{\vec{\xi}}\theta_{AB}&=\xi^i\partial_i\theta_{AB},\\
    \{\Jcal[\theta],\Rcal[\chi]\}&=0,\\
    \{\vec{\Rcal}[\vec{\xi}],\vec{\Rcal}[\vec{\xi}']\}&=-\vec{\Rcal}\big[[\vec{\xi},\vec{\xi}']\big], & [\vec{\xi},\vec{\xi}']^i&=\xi^j\partial_j\xi'^i-\xi'^j\partial_j\xi^i,\\[0.5mm]
    \{\Rcal[\chi],\vec{\Rcal}[\vec{\xi}]\}&=\Rcal[\Lcal_{\vec{\xi}}\chi], & \Lcal_{\vec{\xi}}\chi &= \xi^i \partial_i \chi,\\
    \{\Rcal[\chi],\Rcal[\chi']\}&=\vec{\Rcal}\big[\vec{S}(\chi,\chi')\big], &\quad S^i(\chi,\chi')&=\chi\partial^i \chi' - \chi'\partial^i \chi,
\end{align}
\end{subequations}
for some scalars $\chi, \chi'$, spatial vector fields $\xi^i, \xi'^i$ and antisymmetric $\theta_{AB}$, $\theta_{AB}'$, and $\partial^i=\gamma^{ij}\partial_j$.

\eqref{full_alg} is the local Lorentz-covariant diffeomorphism algebra, from which it is clear that $\Rcal[\chi]$ and $\vec{\Rcal}[\vec{\xi}]$ are the generators of temporal and spatial diffeomorphisms on the canonical variables $E^A_{~i}$ and $\pi^i_A$.\footnote{If we had not written the algebra in terms of $\Rcal$, but instead had used $\widetilde{\Rcal}$, the diffeomorphism algebra would not close, $\{\Rcal(x),\widetilde{\Rcal}_i(y)\}=\Rcal(y)\partial_{y^i}\delta(x-y)-\{\Rcal(x),{}^{(D)}\!\omega_{iAB}(y)\}\Jcal^{AB}(y)$, resulting in field-dependent structure coefficients.}

With all the constraints identified, it is now straightforward to count the dimensionality of the physical phase space by noting that the original phase space, spanned by $E^A_{~i}$, $N$, $N^i$ and their momenta, is $2d^2$ dimensional. Due to the first-class algebra \eqref{full_alg}, all the constraints $\Jcal^{AB}$, $\Rcal,$ and $\Rcal_i$ are first class. It can also easily be verified that the bracket of $P$ and $P_i$ with all other constraints is zero, implying that they are also first-class. Since each of the $d(d+3)/2$ first-class constraints reduces the phase-space dimension by 2, the final phase space has $2d^2-2\times d(d+3)/2=d(d-3)$ dimensions, corresponding to the degrees of freedom of a massless spin-2 field in $d$ dimensions.

\subsection{Castellani Generators}
While $\Rcal$ and $\Rcal_i$ generate diffeomorphisms on the canonical pair $(E^A_{~i}, \pi^i_{A})$, they do not transform the entire phase space. In particular,
\begin{subequations}
\begin{align}
    \{N, \Rcal[\chi]\}&=0, &\{N, \vec{\Rcal}[\vec\xi]\}&=0,\\
    \{N^i, \Rcal[\chi]\}&=0, &\{N^i, \vec{\Rcal}[\vec\xi]\}&=0,    
\end{align}
\end{subequations}
but $N$ and $N^i$ are not scalars under diffeomorphisms. To generalise the generators so that they also transform the lapse and shift and their momenta, we make use of the notion of a Castellani chain \cite{Castellani:1981us}, and derive the full generators,
\begin{subequations}
\label{castellani_GR}
\begin{align}
    G[\chi] &= \Dint x\Big[\chi\Big(\Rcal + \partial^j N\,P_j + \partial_j(PN^j) + \partial_j (NP^j)\!\Big) + \dot{\chi}P\Big],\\
    \vec{G}[\vec\xi] &= \Dint x\Big[\xi^i\Big(\Rcal_i + \partial_i N^jP_j + \partial_j(N^j P_i) + \partial_i NP\Big) + \dot{\xi}^iP_i\Big].
\end{align}
\end{subequations}
These generators can be shown to generate the transformations,
\begin{subequations}
\begin{align}
    \{N,G[\chi]\} &= \dot{\chi}-N^i\partial_i\chi, \\
    \{N^i,G[\chi]\} &= -N\partial^i\chi+\chi\partial^iN, \\
    \{P,G[\chi]\} &= \partial_i(\chi P^i)+P^i\partial_i\chi , \\
    \{P_i,G[\chi]\} &= P\partial_i\chi ,
\end{align}
\end{subequations}
for temporal diffeomorphisms, and 
\begin{subequations}
\begin{align}
    \{N,\vec G[\vec\xi]\} &= \mathcal{L}_{\vec\xi}N = \xi^i\partial_iN, \\
    \{N^i,\vec G[\vec\xi]\} &= \mathcal{L}_{\vec\xi}N^i+\dot{\xi}^i = \xi^k\partial_kN^i-N^k\partial_k\xi^i+\dot{\xi}^i, \\
    \{P,\vec G[\vec\xi]\} &= \mathcal{L}_{\vec\xi}P = \partial_i(\xi^iP), \\
    \{P_i,\vec G[\vec\xi]\} &= \mathcal{L}_{\vec\xi}P_i = \partial_k(\xi^kP_i)+P_k\partial_i\xi^k .
\end{align}
\end{subequations}
for spatial diffeomorphisms, while leaving the transformations of $E^A_{~i}$ and $\pi^i_{A}$ unchanged.

$G[\chi]$ and $\vec G [\vec \xi]$ now generate the gauge transformations on the full phase space, but since they depend on the primary constraints, their brackets closes only weakly,\footnote{The weak identity follows from the fact that one cannot generically construct the gauge generator on the momenta off the constraint surface \cite{Castellani:1981us, Fradkin:1977hw}}
\begin{subequations}
\begin{align}
    \{\vec G[\vec\xi],\vec G[\vec\xi']\}
    &\approx -\vec G\big[[\vec\xi,\vec\xi']\big],
    \\
    \{G[\chi],\vec G[\vec\xi]\}
    &\approx G[\mathcal{L}_{\vec\xi}\chi],
    \\
    \{G[\chi],G[\chi']\}
    &\approx \vec G\big[\vec S(\chi,\chi')\big].
\end{align}
\end{subequations}

\subsection{Lorentz Generator Decomposition}

We conclude this section by decomposing the Lorentz generator $\Jcal^{AB}$ into rotational and boost parts. However, this decomposition cannot be carried out fully without making an explicit choice of the temporal Lorentz vector $X^A$. Nevertheless, we will show that--at the expense of having the resulting algebra close only weakly--we can identify the generators of boosts and rotations.

A generic Lorentz vector $V^A$ can be decomposed into spatial and temporal parts by the use of the projector $P^A_{B}= \delta^A_{B}+ X^A X_B$ \eqref{proj}, 
\begin{align}
\label{vec_dec}
    V^A = \delta^A_B V^B = \Big[P^A_B -X^AX_B \Big]V^B = P^A_BV^B + (-X_BV^B)X^A = V_s^A -V_t X^A
\end{align}
where $V_t = X_BV^B$ and $V_s^A = P^A_B V^B$ are the spatial and temporal projections of $V^A$.

A similar decomposition can be done on the generator $\Jcal^{AB}$ and the function $\theta_{AB}= -\theta_{BA}$ parametrising the Lorentz transformation,
\begin{align}
\label{angle_dec}
    \Jcal^{AB} &= \Jbb^{AB}+ X^{[A} \Kbb^{B]}, &&\theta_{AB} = \vartheta_{AB} + 2X_{[A}\zeta_{B]}, 
\end{align}
where we have defined the generators,
\begin{align}
\label{rot_boost}
    \Jbb^{AB}&= P^A_C P^B_D \Jcal^{CD}, & \Kbb^A &= 2P^A_C X_D \Jcal^{CD},
\end{align}
and the parameters,
\begin{align}
\label{rot_boost_para}
     \vartheta_{AB}&= P^C_A P^D_B \theta_{CD}, & \zeta_A &= P^C_A X^D \theta_{CD}.
\end{align}
With this decomposition, the smeared operator $\Jcal[\theta]$, takes the form,
\begin{subequations}
\begin{align}
\label{J_decom}
    \Jcal[\theta] &= \Jbb[\vartheta] - \Kbb[\vec{\zeta}],\\
\label{rot_gen}
    \Jbb[\vartheta]&= \Dint x \,\vartheta_{AB} \Jbb^{AB},\\
\label{boost_gen}
    \Kbb[\vec{\zeta}]&= \Dint x \,\zeta_A \Kbb^{A},
\end{align}
\end{subequations}
but note that the decomposition is dependent on the $E^A_{~i}$-dependent $X^A$.

Given that a Lorentz vector by definition transforms under infinitesimal Lorentz transformations as,
\begin{align}
    \delta V^A = \theta^A_{~B}V^B,
\end{align}
the decomposition \eqref{rot_boost_para}, yields,
\begin{align}
\label{vec_trans}
    \delta V^A = \vartheta^A_{~B}V^B_s + X^A(\zeta_B V_s^B) - \zeta^AV_t,
\end{align}
where $\vartheta^A_{~B}$ acts only on the spatial part $V_s^A$, while $\zeta_A$ maps a spatial part into a temporal direction $X^A(\zeta_B V_s^B)$ and the temporal projection into a spatial part $- \zeta^AV_t$, as expected.

Applying this to $E^A_{~i}$, and using that $X_A E^A_{~i}$, the spatial vielbein transforms as,
\begin{align}
\label{vielbein_trans}
    \delta E^A_{~i} = \vartheta^A_{~B}E^B_{~i} - X^A(\zeta_B E^B_i),
\end{align}
of which parts directly can be generated by \eqref{rot_gen} and \eqref{boost_gen},
\begin{subequations}
\begin{align}
    \{E^A_{~i}, \Jbb[\vartheta]\}&= \vartheta^A_{~B}E^B_{~i}, \\
    \{E^A_{~i}, \Kbb[\vec{\zeta}]\}&= -X^A(\zeta_B E^B_{~i}).
\end{align}
\end{subequations}
 The vielbein momenta should transform as:
 \begin{align}
\label{momenta_trans}
    \delta \pi^i_A = - \pi^i_B\vartheta^B_{~A} + X_A (\zeta^B\pi^i_B)- \zeta_A(X^B\pi^i_B),
\end{align}
however, since the decomposition \eqref{J_decom} is $X^A$-dependent and $\{X^A, \pi^i_B\}\neq 0$ for generic $X_A$, the generators (\ref{rot_gen}-\ref{boost_gen}) pick up additional terms acting on $\pi^i_A$,
\begin{subequations}
\begin{align}
    \{\pi^i_A,\, \Jbb[\vartheta]\} &=- \pi^i_B \vartheta^B_{~A} + \vartheta_{CD}\{\pi^i_A, P^C_E P^D_F \}\Jcal^{EF}\notag\\
    &\approx  -\pi^i_B \vartheta^B_{~A},\\
    \{\pi^i_A, \Kbb[\vec{\zeta}]\}&= X_A (\zeta^B \pi^i_B)- \zeta_A (X^B \pi^i_B)-2\zeta_B\{\pi^i_A, P^B_C X_D\}\Jcal^{CD}\notag\\
    &\approx X_A (\zeta^B \pi^i_B)- \zeta_A (X^B \pi^i_B).
\end{align}
\end{subequations}
The correct transformation is recovered only on the primary constraint surface $\Jcal^{AB}=0$. This also means that the decomposed Lorentz algebra only holds weakly,
\begin{subequations}
\label{momenta_fail}
\begin{align}
    \{\Jbb[\vartheta],\Jbb[\vartheta']\} &\approx \Jbb\big[[\vartheta,\vartheta'] \big], & [\vartheta, \vartheta']_{AB}&=\vartheta_{AC}\vartheta'^{C}_{~\,B} - \vartheta'_{AC}\vartheta^C_{~B},\\
    \{\Jbb[\vartheta],\Kbb[\vec{\zeta}]\} &\approx \Kbb\big[ \vartheta \cdot \zeta \big], &  (\vartheta \cdot \zeta)_A &= \vartheta_{AB}\zeta^B, \\
    \{\Kbb[\vec{\zeta}],\Kbb[\vec{\zeta}']\} &\approx \Jbb\big[\zeta \wedge \zeta' \big], & (\vec{\zeta}\wedge \vec{\zeta}' )_{AB}& = \zeta_A \zeta'_B - \zeta'_A \zeta_B.
\end{align}
\end{subequations}
Similarly, the brackets between the rotation and boost generators with the Hamiltonian and momentum constraints hold only weakly,
\begin{subequations}
\begin{align}
    \{\Jbb[\vartheta], \Rcal[\chi]\} &\approx 0,\\
    \{\Jbb[\vartheta], \Rcal[\vec{\xi}]\} &\approx \Jbb[\Lcal_{\vec{\xi}}\vartheta]\\
    \{\Kbb[\vec{\zeta}], \Rcal[\chi]\}&\approx0,\\
    \{\Kbb[\vec{\zeta}], \Rcal[\vec{\xi}]\} & \approx \Kbb[\Lcal_{\vec{\xi}}\zeta],
\end{align}
\end{subequations}
as the additional terms in \eqref{momenta_fail} only vanish on the primary constraint surface $\Jcal^{AB}$.

\section{SO(\textit{D}) Covariant Canonical Formulation}
\label{sec:soD_covariant}

In this section, we present an alternative and common canonical parametrisation of the vielbein that manifestly retains the internal $\mathrm{SO}(D)$ invariance by a $D+1$ decomposition of the internal Lorentz space. We then show how to recover the full $\mathrm{SO}(1,D)$ invariance.

\subsection{Internal \texorpdfstring{$D{+}1$}{D+1} split and time gauge}
\label{int_D+1_split}

In Section \ref{sec:Lorentz_covariant} the internal Lorentz space was fully covariant, but it is sometimes convenient to--in analogy to the spacetime--perform a $D+1$ decomposition of the Lorentz space. Using the already formulated vielbein parametrisation \eqref{e_decomp}, this can be done by decomposing the internal unit timelike vector $X^A$ as the $(A,0)$ components of a Lorentz transformation $\Lambda^A_{~B}$,
\begin{align}
\label{X_split}
    X^A=\Lambda^A_{~0}=\begin{pmatrix}\alpha\\ p^a\end{pmatrix},
    \qquad
    \alpha^2 = 1+p^a\delta_{ab}p^b,
    \qquad
    X_A X^A = -\alpha^2 + p^a\delta_{ab}p^b=-1,
\end{align}
where $p^a$ is a boost parameter, $a=1,\ldots,D$ and $\delta_{ab}= \eta_{ab}$ is the flat Euclidean metric. The orthogonality condition $X_AE^A_{~i}=0$ restricts $p^a$ and implies,
\begin{align}
\label{orthogonality_E0i}
    X_AE^A_{~i}=-\alpha E^0_{~i} + p_a E^a_{~i} =0
    \qquad\Longrightarrow\qquad
    p_a E^a_{~i}=\alpha E^0_{~i},
\end{align}
which can be solved explicitly for $p^a$ in terms of the components of $E^A_{~ i}$,
\begin{align}
\label{boost_sol}
    p^a=\frac{ E^0_{~ i} \gamma^{ij}E^a_{~j}}{\sqrt{1-v^2}}, \qquad  v^2=\gamma^{ij}E^0_{~ i} E^0_{~ j}
\end{align}
or alternatively be used to eliminate the components $E^0_{~i}$ in favour of $p^a$. Yet, we will instead introduce $\bar X^A=(1,0,\ldots,0)$ and write,
\begin{align}
\label{X_Lambda_Xhat}
    X^A=\Lambda^A_{~B}\bar X^B,
\end{align}
where we parametrise the Lorentz matrix $\Lambda^A_{~ B}$ by only a boost $p^a$,\footnote{In principle, one could generalise this to include also a rotation, which only appears in the second column and thus not affect the form of $X^A$, but since we will absorb any rotations into the spatial vielbein, this will be redundant.}
\begin{align}
\label{Lambda_boost_rotation}
    \Lambda^A_{~B}
    =
    \begin{pmatrix}
        \alpha & p_b \\
        p^a & A^a_{~b}
    \end{pmatrix},
    \qquad
    A^a_{~b}=\delta^a_b+\frac{1}{1+\alpha}p^a p_b.
\end{align}
Factoring out $\Lambda^A_{~B}$ from the vielbein $e^A_{~\mu}= \Lambda^A_{~B}\bar{e}^B_{~\mu}$, we can introduce the spatial vielbein $\bar{E}^A_{~i}$,
\begin{align}
\label{E_Lambda_Ehat}
    E^A_{~i}=\Lambda^A_{~B}\,\bar{E}^B_{~i},
\end{align}
where the orthogonality condition implies,
\begin{align}
\label{time_gauge_condition}
    X_AE^A_{~i} = \bar X_A \bar{E}^A_{~i} = -\bar{E}^0_{~i} =0.
\end{align}
This reveals that a boost \eqref{boost_sol} corresponds to the Lorentz matrix $(\Lambda^{-1})^{A}_{~B}$ which transforms a general vielbein $e^A_{~ \mu}$ into the lower-triangular vielbein,
\begin{align}
\label{adm_vielbein_hat}
    \bar{e}^A{}_\mu
    =
    \begin{pmatrix}
        N & 0\\
        \bar{E}^a_{~j}N^j & \bar{E}^a_{~i}
    \end{pmatrix},
\end{align}
often referred to as the ADM vielbein. This condition is sometimes imposed directly on $E^A_{~ i}$ by choosing the so-called ``time gauge'', $E^0_{~i}=0$. However, since we will later reintroduce the boost degree of freedom, we will refrain from gauge fixing in this sense.

Note that since the starting vielbein $e^A_{~\mu}$ is arbitrary up to the definition $g_{\mu \nu}=e^A_{~\mu}\eta_{AB}e^B_{~\nu}$, leaving the Lorentz frame undetermined, the boost $p^a$ is fully generic and only parametrises the $D$ fields $E^0_{~ i} = p_aE^a_{~ i}/\alpha$, and $E^A_{~i} \to(\bar{E}^a_{~i}, p^a)$ does not change the number of fields.  The full general vielbein decomposition reads,
\begin{align}
\label{e_boost}
    e^A{}_\mu = \Lambda^A_{~B}\,\bar{e}^B{}_\mu = 
    \begin{pmatrix}
        \alpha & p_b \\
        p^a & A^a_{~b}
    \end{pmatrix}\begin{pmatrix}
        N & 0\\
        \bar{E}^b_{~j}N^j & \bar{E}^b_{~i}
    \end{pmatrix} = \begin{pmatrix}
        \alpha N +  p_b\bar{E}^b_{~j}N^j & p_b\bar{E}^b_{~i}\\
        N p^a +  A^a_{~b}\bar{E}^b_{~j}N^j & A^a_{~b}\bar{E}^b_{~i}
    \end{pmatrix}.
\end{align}
In this parametrisation, the spatial metric is constructed entirely in terms of the spatial vielbein $\bar{E}^a_{~ i}$,
\begin{align}
\label{gamma_from_E}
    \gamma_{ij}=g_{ij}= \bar{e}^C_{~ i}\Lambda^A_{~ C}\eta_{AB}\Lambda^B_{~D}\bar{e}^D_{~ j}=
    \bar{E}^a_{~i}\,\delta_{ab}\,\bar{E}^b_{~j},
\end{align}
which is manifestly invariant under $\mathrm{SO}(D)$ rotations of the vielbein $\bar{E}^a_{~i}\mapsto \Omega^a_{~b}\,\bar{E}^b_{~i}$, but not invariant under a boost, since this does not preserve the lower-triangular form \eqref{adm_vielbein_hat}.

We can define the inverse vielbein, $\bar{E}{}^i_{~ a} = \gamma^{ij}\delta_{ab}\bar{E}{}^b_{~ j}$, which in contrast to $E^A_{~ i}$, fulfils being both the spatial and SO($D$) inverse,
\begin{align}
    \bar{E}{}^a_{~ i}\bar{E}{}^i_{~ b}=\delta^a_b, && \bar{E}{}^i_{~ a}\bar{E}{}^a_{~ j}= \delta^i_j,
\end{align}
and its components are given by the matrix inverse $(\bar{E}{}^{-1})^i_{~ a}$ of the invertible $D \times D$ matrix $\bar{E}{}^a_{~ i}$. $\bar{E}^i_{~a}$ can be used to write the inverse vielbein $\bar{e}^\mu_{~A}$ in lower-triangular form,
\begin{align}
    \bar{e}^\mu_{~A} = \frac{1}{N}\begin{pmatrix}
        1 & 0 \\
        -N^i & N\bar{E}^i_{~a}
    \end{pmatrix}.
\end{align}

\subsection{SO(\textit{D}) invariant phase-space action}
\label{sec:so_phase_space_action}

In this section we will derive the $\mathrm{SO}(D)$ invariant phase-space form of the Einstein–Hilbert action using the vielbein parametrisation of the previous section. The derivation follows the same logic as for the Lorentz-covariant form in Section \ref{sec:Lorentz_covariant}, but for completeness, we will provide a self-contained section. From this point onwards, we suppress the bar notation on the vielbein \eqref{adm_vielbein_hat} for notational convenience and simply write $E^{a}_{~ i}$. Likewise, we will introduce new curvatures, constraints, covariant derivatives, spin connections, and related quantities, so rather than cluttering the notation with bars throughout, we will reuse the notation from the previous section. We therefore advise \textbf{caution} against mixing equations across the different Sections without properly considering what will change.

We will start with the Einstein–Hilbert action from Section \ref{sec:D+1_EH}, written in terms of the spatial and extrinsic curvatures $\rD \!R$ and $K_{ij}$, where we have disregarded the boundary term $\rd\nabla_\mu v^\mu$ \eqref{Gauss-Codazzi},
\begin{align}
    \Scal= \mpl \dint x\, N\sqrt{\gamma}\Big [\!\rD \!R+ K^{ij}K_{ij}-K^2\Big]. 
\end{align}
The only time-derivatives enter the action through the extrinsic curvature via the spatial metric \eqref{gamma_from_E},
\begin{align}
\label{gamma_dot_2}
    \dot\gamma_{ij}&=2\delta_{ab}E^a_{~(i}\dot E^b_{~j)},\\
    K_{ij}&= \frac{1}{N}\Big[E^{\,}_{a(i}\dot{E}^a_{~ j)}-\rD \nabla_{(i} N_{j)}\Big].
\end{align}
Since $E^a_{~ i}$ only contains the spatial metric and rotational degrees of freedom, we emphasise that the action is independent of any time derivatives of the boosts in this formalism. In fact, the Lorentz invariance of the action leaves the action manifestly independent of the boosts $p^a$, so that it can be ignored in the canonical analysis.

The momenta conjugate to the SO($D$) covariant spatial vielbein $E^a_{~i}$ takes the form,
\begin{align}
\label{pi_soD}
    \pi^i_{~a} =\frac{\delta\Scal }{\delta \dot E^a_{~i}}
    =2\mpl\sqrt{\gamma}E_{aj}\Big[K^{ij}-\gamma^{ij}K\Big].
\end{align}
Since the rotational degrees of freedom are not dynamical, the $D^2$ components of $\pi^i_{~a}$ are not all independent. It follows from the symmetry of $K^{ij}$ and $\gamma^{ij}$, that $\pi^{i[a}_{\phantom{i}}E^{b]}_{~i}$ vanish identically from the definition of the momenta, generating $D(D-1)/2$ primary constraints,
\begin{align}
\label{Jab}
    \Jcal^{ab}=\pi^{i[a}_{\phantom{i}}E^{b]}_{~i} \approx 0.
\end{align}
This reflects that the Einstein--Hilbert action only depends on the rotation independent symmetric combination \eqref{gamma_dot_2}, and constrains the $D^2$ components of $\pi^i_{~ a}$ to $D(D+1)/2$ independent fields. The anti-symmetric part $\dot{w}_{ij}=2\delta_{ab}E^a{}_{[i}\dot E{}^b{}_{j]}$ instead correlates to the rotational velocities, so that,
\begin{align}
    E_{ai}\dot{E}^a_{~ j}= E^{}_{a(i}\dot{E}^a_{~ j)}+E^{}_{a[i}\dot{E}^a_{~ j]} = \tfrac{1}{2}(\dot{\gamma}_{ij} + \dot{w}_{ij}).
\end{align}
The momenta conjugate to $w_{ij}$ is proportional to $\Jcal^{ab}$,
\begin{align}
    \frac{\delta \Scal}{\delta \dot{w}_{ij}}= \frac{\delta \Scal}{\delta \dot{E}^a_{~ k}} \frac{\delta \dot{E}^a_{~ k}}{\delta \dot{w}_{ij}}= \tfrac{1}{2}\pi^k_{~a}E^{al}\delta^{[i}_l\delta^{j]}_k=-\tfrac{1}{2} E^i_{~ a}E^j_{~ b}\Jcal^{ab},
\end{align}
showing that the primary constraint \eqref{Jab} are precisely the momenta of the rotational fields expressed in the Lorentz frame.

The presence of the primary constraints $\Jcal^{ab}=0$ means that we cannot invert \eqref{pi_soD} for the velocities $\dot{E}^a_{~ i}$ uniquely, but we would need to resort to the notion of the pseudoinverse (see Appendix \ref{pseudo_inv_method}). We can, however, invert \eqref{pi_soD} for the extrinsic curvature off the primary constraint surface $\Jcal^{ab}=0$,
\begin{align}
\label{Kij_soD}
    \Kcal^{i}{}_{j}
    &=\frac{1}{2\mpl\sqrt{\gamma}}
    \Big[\pi^{i}{}_{a}E^{a}{}_{j}-\tfrac{1}{d-2}\delta^{i}_{j}\pi^{k}{}_{a}E^{a}{}_{k}\Big].
\end{align}
Where only the symmetric part $\Kcal_{(ij)}= K_{ij}(E, \pi)$ appears in the action, and the anti-symmetric part is proportional to the primary constraints,
\begin{align}
    \Kcal_{[ij]} = \frac{-1}{2\mpl\sqrt{\gamma}}E_{ai}E_{bj}\Jcal^{ab},
\end{align}
so that $\Kcal_{ij} \approx K_{ij}$. 

To obtain the phase-space form, we proceed with a similar trick as in Section \ref{sec:Phase_space_Action} and evaluate the Lorentz-covariant time-derivative of $\bar{e}^A_{~\mu}$ \eqref{adm_vielbein_hat},\footnote{We could instead define the time derivative of $e^{A}_{~\mu}$ (including the boost components), but this would make the spin connection slightly more complicated by introducing time derivatives of the boosts. These terms would, however, be proportional to the primary constraint that the boost momenta vanish.}
\begin{align}
\label{cov_derivative_e_bar_soD}
    \rd\Dcal_\mu \bar{e}^A{}_\nu
    =\partial_\mu \bar{e}^A{}_\nu - \rd\!\Gamma^\sigma_{~\mu\nu}\bar{e}^A{}_\sigma+\rd\!\omega_\mu{}^{A}{}_{B}\,\bar{e}^B{}_\nu =0,
\end{align}
where $\rd\Gamma^\sigma_{~\mu \nu}$ are the Christoffel symbols of the $g_{\mu \nu}$-compatible covariant derivative $\rd \nabla_\mu$,  $\rd\nabla_\mu g_{\alpha \beta}=0$ and $\rd \omega_{\mu AB} = \bar{e}^\sigma_{~[A}\rd\nabla_\mu \bar{e}^{\phantom{A}}_{B] \sigma}$ is the spin-connection.

Restricting \eqref{cov_derivative_e_bar_soD} to $(\mu,\nu,A)=(0,i,a)$ yields the covariant time derivative of $\bar{e}^a_{~ i}=E^a_{~ i}$. Using the standard $D+1$ identity (see Appendix \ref{sec:Christoffel}),
\begin{align}
    \rd\!\Gamma^0{}_{0i}N^j+\rd\!\Gamma^j{}_{0i}=NK^j_{~i}+\rD\nabla_i N^j,    
\end{align}
where $\rD \nabla_i$ is the $\gamma_{ij}$-compatible covariant derivative, \eqref{cov_derivative_e_bar_soD} reads,
\begin{align}
    \rd\Dcal_0 \bar{e}^a_{~ i}=\dot{E}^a_{~ i} -N K^{j}{}_{i}E^a{}_{j}-E^a{}_{j} \rD\nabla_i N^j+ \rd\!\omega_0{}^a_{~b}E^b_{~ i}=0,
\end{align}
which is manifestly $\mathrm{SO}(D)$ covariant, since $\dot{E}^a_{~ i}+ \rd\!\omega_0{}^a_{~b} E^b_{~ i}$ transforms covariantly. If we contract with the unconstrained $\pi^i_{~a}$, and use the $D+1$ form of the spin-connection \eqref{time_gauge_w}, we obtain,
\begin{align}
\label{piEdot_soD}
    \pi^i_{~a}\dot E^a_{~i} 
    -N K^{j}{}_{i}E^a{}_{j}\pi^i_{~a} 
    -E^a{}_{j}\pi^i_{~a} \rD\nabla_i N^j
    + \Big[E^i_{~[a}\dot{E}^{\phantom{i}}_{b]i}-E^i_{~[a}E^j_{~b]}\rD\nabla_i N_j \Big]\Jcal^{ab}=0.
\end{align}
If we use \eqref{Kij_soD} and $K_{ij}= \Kcal_{(ij)}(E, \pi)$ to express the Einstein–Hilbert Lagrangian in terms of $\pi^i_{~a}$, it can be written in the form,
\begin{align}
    \Lcal=N\mpl \sqrt{\gamma}\Big[\!\rD \!R+K_{ij}K^{ij}-K^2 \Big] =NK^j_{~ i}\pi^i_{~a}E^a_{~j}+ N\mpl \sqrt{\gamma}\Big[\!\rD \! R-K_{ij}K^{ij}+K^2 \Big],
\end{align}
where the term $NK^j_{~ i}\pi^i_{A}E^A_{~j}$ is common with the second term in \eqref{piEdot_soD}, and a substitution yields the phase-space Lagrangian,
\begin{align}
\label{so3_lagrangian}
    \Lcal&= \pi^i_{~a} \dot{E}^a_{~ i}+N\mpl \sqrt{\gamma}\Big[\!\rD \! R-\Kcal_{ij}\Kcal^{ij}+\Kcal^2 \Big]-\pi^i_{~a}E^a_{~j}\rD\nabla_i N^j + W_{ab}\Jcal^{ab},\\[2mm]
\label{Omega_def_so3}
    W_{ab}&=E^i_{[a}\dot{E}^{}_{b] i}- E^i_{[a}E^j_{b]}\rD\nabla_i N_j -\tfrac{N}{2}E^i_{[a}E^j_{b]}\Kcal_{ij}.
\end{align}
Here we have written the lapse proportional term of \eqref{so3_lagrangian} in terms of $\Kcal_{ij}= K_{ij}+\Kcal_{[ij]}$, and factored the pure anti-symmetric part $\Kcal_{[ij]}$ into $W_{ab}\Jcal^{ab}$.

Substituting this into the action, and partially integrating the term $\pi^i_{~a}E^a_{~j}\rD\nabla_i N^j$, we obtain the canonical phase-space action,
\begin{align}
\label{EH_phase_space_soD}
    \Scal
    =\dint x\,
    \Big[
        \pi^i_{~a}\dot E^a_{~i}
        +N\Rcal
        +N^i\widetilde{\Rcal}_i
        +W_{ab}\Jcal^{ab}
    \Big],
\end{align}
where $\Rcal$ and $\widetilde{\Rcal}_i$ are defined by,
\begin{subequations}
\label{R_mu_soD}
\begin{align}
\label{R_constraint_soD}
    \Rcal
    &=
    \mpl\sqrt{\gamma}\rD \! R
    +\tfrac{1}{4\mpl\sqrt{\gamma}}
    \left[
        \tfrac{1}{d-2}\big(\pi^i_{~a} E^a_{~ i}\big)^2
        -\pi^i_{~ a} E^a_{~ j}\pi^j_{~ b} E^b_{~ i}
    \right],
    \\
\label{R_i_constraint_soD}
    \widetilde{\Rcal}_i
    &= E^a_{~i}\rD\Dcal_j\pi^j_{~a} = E^a_{~ i}\Big[ \partial_j \pi^j_{~ a}- \rD\!\omega_{ja}{}^b \pi^j_{~ b}\Big],
\end{align}
\end{subequations}
where $\rD\Dcal_i$ is the $\mathrm{SO}(D)$-covariant derivative compatible with $E^a_{~i}$, $\rD \Dcal_i E^a_{~ j}=0$, with corresponding Christoffel symbols $\rD\Gamma^i_{~ jk}$ and spin-connection $\rD\!\omega_{iab}=E^j{}_{[a}\rD\nabla_i E_{b]j}$.

We could in principle make a comparison of this form of the action with the metric form $\int \mathrm{d}^dx[\pi^{ij}\dot{\gamma}_{ij}+ N\Rcal^\gamma+N^i \Rcal^\gamma_i ]$, but the comparison would be identical to the one in Appendix \ref{sec:metric_action}, but with $\pi^i_{~ 0}=E^0_{~ i}= X^a=0$. We therefore omit the repitition for brevity, but stress that the additional terms come from the relation between the metric canonical one-form and form of the constraints in the vielbein formulation, generating additional terms $W_{ab}\Jcal^{ab}$. At the level of the action, the metric and SO($D$) covariant form are identical, but the functions \eqref{R_mu_soD} act as generators also in the rotational degrees of freedom, in contrast to $\Rcal^\gamma$ and $\Rcal^\gamma_i$ which only transform $\gamma_{ij}$ and $\pi^{ij}= \tfrac{1}{2}\pi^{(i}_{~A}E^{j)A}_{\phantom{A}}$ (see longer discussion in Appendix \ref{sec:metric_action}).

\subsection{Hamiltonian formulation}

We can now construct the total Hamiltonian from the phase-space action \eqref{EH_phase_space_soD}. Introducing the momenta conjugate to the lapse and shift, 
\begin{align}
    P = \frac{\delta \Scal}{\delta \dot{N}}, &&     P_i = \frac{\delta \Scal}{\delta \dot{N}^i}
\end{align}
we note that, since we have removed the boundary term $\rd \nabla_\mu v^\mu$ \eqref{Gauss-Codazzi} from the action, these derivatives are zero and produce the trivial primary constraints $P\approx0$ and $P_i\approx0$. Adding the primary constraints $P\approx 0, P_i \approx0$ and $\Jcal^{ab}\approx0$ with Lagrange multipliers, the total Hamiltonian takes the form,
\begin{align}
    H_T
    =-\Dint x
    \Big[
        N\Rcal + N^i \widetilde{\Rcal}_i
        +\lambda P + \lambda^i P_i
        +\widetilde{\lambda}_{ab}\Jcal^{ab}
    \Big].
\end{align}
Here we have absorbed $W_{ab}$ into the arbitrary Lagrange multiplier $\widetilde{\lambda}_{ab}$. 

The function $\widetilde{\Rcal}_i$ is, however, not the generator of pure spatial diffeomorphisms. In Appendix \ref{sec:mod_gen_1} we derive the Lorentz-covariant generator of spatial diffeomorphisms, and show that the analogous function $\widetilde{\Rcal}_i$ needs to be modified. A similar argument implies that the generator differs from $\widetilde{\Rcal}_i$ with a term proportional to $\Jcal^{ab}$,
\begin{align}
    \Rcal_i = \widetilde{\Rcal}_i+ \rD \!\omega_{iab}\Jcal^{ab},
\end{align}
which can be implemented into the Hamiltonian by a shift of the Lagrange multiplier $\widetilde{\lambda}_{ab}$,
\begin{align}
    \lambda_{ab} = \widetilde{\lambda}_{ab}- N^i\rD\!\omega_{iab},
\end{align}
to yield the final expression for the total Hamiltonian,
\begin{align}
\label{HT_soD}
    H_T
    =-\Dint x
    \Big[
        N\Rcal + N^i \Rcal_i
        +\lambda P + \lambda^i P_i
        +\lambda_{ab}\Jcal^{ab}
    \Big].
\end{align}
With the variables $(N, P),\, (N^i, P_j), \,(E^a_{~ i}, \pi^j_b)$, we can introduce the canonical Poisson brackets,
\vspace{-5mm}
\begin{subequations}
\label{PB_soD}
\begin{align}
    \{N(x),P(y)\}&=\delta(x-y),\\
    \{N^i(x),P_j(y)\}&=\delta^i_j\delta(x-y),\\
    \{E^a_{~i}(x),\pi^j_{~b}(y)\}&=\delta^a_b\delta^j_i\delta(x-y),    
\end{align}    
\end{subequations}
where all other brackets between the canonical variables vanish identically.

Hamilton's equations now yield the field equations for the vielbein and its momentum,
\begin{subequations}
\label{Ham_eq_SO(D)}
\begin{align}
\label{Edot_soD}
    \dot{E}^a_{~i} &\approx \{E^a_{~i}, H_T\} \approx NK^j_{~i}E^a_{~j} + \Lcal_{\vec{N}} E^a_{~i}+ \lambda^a_{~b}E^b_{~i},\\
    \dot{\pi}^{i}_{~a} &\approx \{ \pi^{i}_{~a}, H_T\} \approx \frac{\delta \Rcal[N]}{\delta E^a_{~i}}+ \Lcal_{\vec{N}} \pi^{i}_{~a}+\lambda_a^{~b}\pi^i_{~b} ,\\
\label{N_dot}
    \dot{N}&\approx \{N, H_T\} \approx - \lambda, \\
\label{N_idot}
    \dot{N}^i&\approx \{N^i, H_T\} \approx - \lambda^i,
\end{align}
\end{subequations}
where we have used the Lie derivative form of $E^a_{~i}$ and $\pi^i_{~a}$ \eqref{diff} and,
\begin{align}
     \frac{\delta \Rcal[N]}{\delta E^a_{~i}}&=-2\mpl \sqrt{\gamma} E_{aj}\Big[N\Big(\rD \!R^{ij}-\tfrac{1}{2}\gamma^{ij}\rD \!R\Big)+\gamma^{ij}\rD\nabla^2 N-\rD\nabla^i\rD\nabla^j N\Big]\\
    &\quad\,+\tfrac{N}{4\mpl \sqrt\gamma}\Big[\tfrac{2}{d-2}\big(\pi^k_c E^c_{~k}\big)\pi^i_{~a}-2\pi^i_{~b} E^b_{~j}\pi^j_{~a}-\Big(\tfrac{1}{d-2}\big(\pi^k_{~c} E^c_{~ k}\big)^2-\pi^m_{~b} E^b_{~ n}\pi^n_{~c} E^c_{~m}\Big)E^i_{~a}\Big].\notag
\end{align}
The equations of motion for the lapse and shifts (\ref{N_dot}-\ref{N_idot}) determine the Lagrange multipliers $\lambda $ and $\lambda^i$ after gauge fixing, while we can anti-symmetrise \eqref{Edot_soD} to determine $\lambda_{ab}$,
\begin{align}
\label{lambda_sol}
    \dot{E}^{}_{[ai}E^i_{~ b]} &\approx \{E_{[ai}, H_T\}E^i_{~ b]} \approx E^i_{[a}E^j_{~b]}\rD\nabla_j N_i- \tfrac{N}{2}E^i_{~[a}E^j_{~b]}\Kcal_{ij}- N^j \rD\!\omega_{jab} - \lambda_{ab},
\end{align}
where, apart from $\lambda_{ab}$, we re-obtain the terms $W_{ab}$ and $N^j \rD\!\omega_{jab}$ previously absorbed into it.

\subsection{Constraints and Symmetry Operators}
We will now follow the Dirac–Bergmann algorithm and obtain the constraints and their first-class algebra from the Hamiltonian \eqref{HT_soD}. 

Imposing the stability of the momenta conjugate to the lapse and shift, we obtain the vielbein Hamiltonian and momentum constraints,
\begin{subequations}
\begin{align}
\label{ham_const_1}
    \dot{P}(x) &= \{P(x),\, H_T\}\approx \Rcal\,(x) \,\approx0,\\
\label{mom_const_2}
    \dot{P}_i(x) &= \{P_i(x), H_T\}\approx \Rcal_i(x) \approx0,
\end{align}
\end{subequations}
while imposing $\dot{\Jcal}^{ab}\approx 0$, results in the condition,
\begin{align}
    \{\Jcal^{ab}(x), H_T\} 
    \approx 
    -\Dint y \Big[N(y)\{\Jcal^{ab}(x),\Rcal(y)\}
        &+N^i(y)\{\Jcal^{ab}(x),\Rcal_i(y)\}\notag\\
\label{Jdot_so3}
        &+\lambda_{cd}(y)\{\Jcal^{ab}(x),\Jcal^{cd}(y)\}
    \Big] \approx0.
\end{align}
If we introduce the smeared operator,
\begin{align}
\label{smeared_so3}
    \Jcal[\vartheta] = \Dint x\, \vartheta_{ab}(x)\Jcal^{ab}(x), && \vartheta_{ab}=-\vartheta_{ba},
\end{align}
a straightforward computation shows that $\Jcal[\vartheta]$ generates infinitesimal rotations $\Omega^a_{~ b} = \delta + \vartheta^a_{~ b}$ on the canonical vielbein variables,
\begin{align}
\label{E_pi_SO(D)_trans}
    \{E^a_{~ i}(x), \Jcal[\vartheta]\}= \vartheta^a_{~ b}(x)E^b_{~ i}(x), &&
    \{\pi^i_{~ a}(x), \Jcal[\vartheta]\}= -\pi^i_{~ b}(x)\vartheta^b_{~ a}(x).
\end{align}
Using this, the brackets in \eqref{Jdot_so3} can be computed,
\begin{subequations}
\label{J_brackets_so3}
\begin{align}
\label{so3_lie_algebra}
    \{\Jcal^{ab}(x),\,\Jcal^{cd}(y)\}
    &=\Big[\delta^{a[c}\,\Jcal^{d]b}(x)-\delta^{b[c}\,\Jcal^{d]a}(x)\Big]\,\delta(x-y),\\
\label{JR_i_so3}
    \{\Jcal^{ab}(x),\Rcal_i(y)\}
    &= \Jcal^{ab}(y)\, \pdv{}{y^i}\delta(x-y),\\
\label{JR_so3}
    \{\Jcal^{ab}(x), \Rcal(y)\}
    &=0,
\end{align}
\end{subequations}
where \eqref{so3_lie_algebra} is the SO($D$) algebra, which if $D=3$ can be written in the more familiar form,
\begin{align}
    \{\Jcal_a(x),\Jcal_b(y) \} = \epsilon_{ab}{}^c \Jcal_c(x) \delta(x-y), && \Jcal_a = -\tfrac{1}{2}\epsilon_{abc}\Jcal^{bc}.
\end{align}
Since all the brackets in \eqref{Jdot_so3} vanish weakly by \eqref{J_brackets_so3}, $\dot{\Jcal}^{ab} \approx 0$ does not impose any secondary constraints. 

Imposing the stability of the secondary constraints (\ref{ham_const_1}–\eqref{mom_const_2} yields,
\begin{subequations}
\label{RRi_dot_so3}
\begin{align}
\label{R_dot_so3}
    \!\!\!\dot{\Rcal}(x) &\approx \!-\Dint y\Big[N(y) \{\Rcal(x)\;,\Rcal(y)\}+N^j(y) \{\Rcal(x)\;,\Rcal_j(y)\} + \lambda_{ab}(y) \{\Rcal(x)\;,\Jcal^{ab}(y)\} \Big]\approx0,\\
\label{Ri_dot_so3}
    \!\!\!\dot{\Rcal}_i(x) &\approx\! -\Dint y\Big[N(y) \{\Rcal_i(x),\Rcal(y)\}+N^j(y) \{\Rcal_i(x),\Rcal_j(y)\} + \lambda_{ab}(y) \{\Rcal_i(x),\Jcal^{ab}(y)\} \Big]\approx0,
\end{align}
\end{subequations}
where the last brackets in each line vanish by \eqref{J_brackets_so3}. The remaining brackets form the first-class diffeomorphism algebra, which can be computed in direct analogy to the metric formulation (see \cite{Khoury:2011ay} Appendix A),
\begin{subequations}
\label{diff_alg_so3}
\begin{align}
\label{RR_2}
     \{\Rcal(x),\Rcal(y)\}
     &= \left[\Rcal^i(y)\pdv{}{y^i}-\Rcal^i(x)\pdv{}{x^i}\right]\delta(x-y),\\
\label{RR_i_2}
     \{\Rcal(x),\Rcal_i(y)\}
     &=\Rcal(y)\pdv{}{y^i}\delta(x-y),\\
\label{R_iR_j_2}
    \{\Rcal_i(x), \Rcal_j(y)\}
    &=\left[\Rcal_i(x) \pdv{}{y^j}-\Rcal_j(y)\pdv{}{x^i}\right]\delta(x-y),
\end{align}
\end{subequations}
where $\Rcal^i= \gamma^{ij}\Rcal_j$. The smeared functions,
\begin{align}
\label{smeared_R}
    \Rcal[\chi] = \Dint x\, \chi(x) \Rcal(x), && \vec{\Rcal}[\vec{\xi}] = \Dint x\, \xi^i(x) \Rcal_i(x),
\end{align}
are generators of temporal and spatial diffeomorphisms parametrised by a scalar $\chi$ and spatial vector field $\xi^i$.

Using the smeared operators \eqref{smeared_so3} and \eqref{smeared_R}, the full constraint algebra takes the form,
\begin{subequations}
\label{full_alg_so3}
\begin{align}
    \{ \Jcal[\vartheta],\Jcal[\vartheta']\}&=\Jcal\big[[\vartheta,\vartheta']\big], 
    & [\vartheta,\vartheta']_{ab}&=\vartheta_{a}^{c}\vartheta'_{cb}-\vartheta'_a{}^{c}\vartheta_{cb},\\
\label{JRi_bracket}
    \{\Jcal[\vartheta],\vec{\Rcal}[\vec{\xi}]\}&=\Jcal[\Lcal_{\vec{\xi}}\vartheta], & \Lcal_{\vec{\xi}}\vartheta_{ab}&=\xi^i\partial_i\vartheta_{ab},\\
\label{JR_bracket}
    \{\Jcal[\vartheta],\Rcal[\chi]\}&=0,\\
    \{\vec{\Rcal}[\vec{\xi}],\vec{\Rcal}[\vec{\chi}]\}&=-\vec{\Rcal}\big[[\vec{\xi},\vec{\chi}]\big], & [\vec{\xi},\vec{\chi}]^i&=\xi^j\partial_j\chi^i-\chi^j\partial_j\xi^i,\\[0.5mm]
    \{\Rcal[\chi],\vec{\Rcal}[\vec{\xi}]\}&=\Rcal[\Lcal_{\vec{\xi}}\chi], & \Lcal_{\vec{\xi}}\chi &= \xi^i \partial_i \chi,\\
    \{\Rcal[\chi],\Rcal[\chi']\}&=\vec{\Rcal}\big[\vec{S}(\chi,\chi')\big], &\quad S^i(\chi,\chi')&=\chi\partial^i \chi' - \chi'\partial^i\chi,
\end{align}
\end{subequations}
which is the SO(\textit{D}) covariant diffeomorphism algebra and we have used the notation  $\partial^i=\gamma^{ij}\partial_j$.

From the first-class algebra it follows that all brackets in \eqref{RRi_dot_so3} are weakly zero, and no tertiary constraints appear. Since the brackets of $P$ and $P_i$ vanish trivially with all the constraints, they are also first-class, resulting in a total of $d(d+3)/2$ first-class constraints, which all reduce the phase-space dimension by 2. Since the initial phase space, spanned by $E^a_{~i}, N, N^i$ and their momenta, is $2d^2-2(d-1)$ dimensional, the resulting physical phase space is $d(d-3)$ dimensional, corresponding to the $d(d-3)/2$ helicities of a massless spin-2 field.

\subsection{Extending the Lorentz Algebra}
In this section, we recover the full Lorentz symmetry of the SO($D$) covariant formulation by extending the phase space and constructing the boost generator. We also provide the modified Lorentz covariant diffeomorphism algebra.

In contrast to explicitly fixing the time gauge in Section \ref{int_D+1_split}, we rewrote the general vielbein in terms of a boosted lower-triangular vielbein,
\begin{align}
\label{full_e}
    e^A{}_\mu = 
    \begin{pmatrix}
        \alpha & p_b \\
        p^a & A^a{}_b
    \end{pmatrix}\begin{pmatrix}
        N & 0\\
        E^b_{~j}N^j & E^b_{~i}
    \end{pmatrix} = \begin{pmatrix}
        \alpha N +  p_b E{}^b_{~j}N^j & p_b E{}^b_{~i}\\
        N p^a +  A^a{}_b E{}^b_{~j}N^j & A^a{}_b E{}^b_{~i}
    \end{pmatrix},
\end{align}
and noted that the boost transformation dropped out of the action. However, one could instead have chosen to retain $p^a$ as a canonical variable. If we extend the phase space by introducing its conjugate momenta,
\begin{align}
    \pi_a = \frac{\delta \Scal }{\delta \dot{p}^a},
\end{align}
we obtain the primary constraint $\pi_a =0$, as the action is manifestly independent of both $p^a$ and $\dot{p}^a$. Note that this extension is trivial in the sense that $\pi_a \approx 0$ are $D$ first-class constraints, thus reducing the dimensionality of the physical phase space by $2D$, leaving the physical dimensionality unchanged. Nevertheless, the introduction of the additional phase space variables $p^a$ and $\pi_a$ allows us to recover the full Lorentz symmetry. 

For an infinitesimal Lorentz transformation $\Lambda^A_{~B} = \delta^A_B + \theta^A_{~B}$ parametrised by $\theta_{AB}= -\theta_{BA}$, the vielbein $e^A_{~\mu}$ transforms as a Lorentz vector,
\begin{align}
\label{e_trans}
    \delta e^A_{~\mu} = \theta^A_{~B}e^B_{~\mu}.
\end{align}
For spatial rotations, we have
\begin{align}
    \vartheta_{AB} = \begin{pmatrix}
        0 & 0 \\
        0 & \vartheta_{ab}
    \end{pmatrix},
\end{align}
but the rotational generator $\Jcal^{ab}=\pi^{i[a}E^{b]}_i$ \eqref{smeared_so3} does not produce this transformation on $e^A_{~\mu}$. It does so only on $E^a_{~i}$, due to the presence of the boost $p^a$, which is needed to parametrise a general vielbein. However, we can extend the rotational generator to transform $p^a$ and its momenta, so that the full vielbein $ e^A_{~\mu}$ transforms correctly \eqref{e_trans}. The extension can easily be shown to be,
\begin{align}
    \Jbb^{ab} = \Jcal^{ab}+ \pi^{[a}p^{b]}.
\end{align}
The additional term does not affect the transformation of $E^a_{~i}$ and $\pi^i_{~a}$ \eqref{E_pi_SO(D)_trans}, but yields,
\begin{align}
    \{p^a, \Jbb[\vartheta] \} = \vartheta^a_{~b}p^b, && \{\pi_a, \Jbb[\vartheta] \} = -\pi_b\vartheta^b_{~a}.
\end{align}
With this, it is easy to show that objects like $A^a_{~b}E^b_{~i}$, $p_a E^a_{~i}$ and $\alpha$ transform tensorially under infinitesimal rotations, so that the components of $e^A_{~\mu}$ transform as,
\begin{subequations}
\begin{align}
    \delta e^0_{~0} &= \{\alpha N + p_b E^b_{~j}N^j, \Jbb[\vartheta] \}=0, &
    \delta e^0_{~i} &= \{ p_b E^b_{~i}, \Jbb[\vartheta] \}=0\\
    \delta e^a_{~0} &= \{Np^a + A^a_{~b} E^b_{~j}N^j, \Jbb[\vartheta] \}=\vartheta^a_{~b}e^b_{~0}, &
    \delta e^a_{~i} &= \{ A^a_{~b} E^b_{~i}, \Jbb[\vartheta] \}=\vartheta^a_{~b}e^b_{~i},
\end{align}
\end{subequations}
implying that $\delta e^A_{~\mu}= \vartheta^A_{~B}e^B_{~\mu}$, as needed.

The tensorial transformation simply yields,
\begin{align}
    \{\Jbb[\vartheta], \Jbb[\vartheta']\} = \Jbb\big[[\vartheta, \vartheta']\big],
\end{align}
where $[\vartheta, \vartheta']_{ab} = \vartheta_{ac}\vartheta'^c_{~b}-\vartheta'_{ac}\vartheta^c_{~b}$, or equivalently,
\begin{align}
    \{\Jbb^{ab}(x),\Jbb^{cd}(y)\}
    &=\Big[\delta^{a[c}\Jbb^{d]b}(x)-\delta^{b[c}\Jbb^{d]a}(x)\Big]\delta(x-y).
\end{align}
Since $\Jcal^{ab}$ and $\pi_a$ are first class, so is $\Jbb^{ab}$, and the above forms a first-class algebra. 

Let us now instead consider the transformation \eqref{e_trans}, but $\theta_{ab}=0$ and $\theta_{a0}= \zeta_a$, parametrising a boost. From \eqref{e_trans} the components transform as,
\begin{subequations}
    \begin{align}
    \label{delta_e0mu}
        \delta e^0_{~0} &= \zeta_b e^b_{0}, &
        \delta e^0_{~i} &= \zeta_b e^b_{~i}\\
    \label{delta_eamu}
        \delta e^a_{~0} &= \zeta^a e^0_{~0}, &
        \delta e^a_{~i} &= \zeta^a e^0_{~i}.
    \end{align}
\end{subequations}
Using the component form of $e^A_{~\mu}$ \eqref{full_e} yields the transformations,
\begin{subequations}
\begin{align}
\label{delta_p}
    \delta p^a &= \alpha \zeta^a, &
    \delta \alpha &=\zeta_a p^a, \\
\label{delta_AE}
    \delta (p_a E^a_{~j}) &=\zeta_a A^a_{~b}E^b_{~j},&
    \delta (A^a_{~b}E^b_{~j})&= \zeta^a p_b E^b_{~j},
\end{align}
\end{subequations}
where the first line determines how $p^a$ must transform, which in turn imposes the transformation behaviour of $A^a_{~b}$. Combining these, we can derive that,
\begin{align}
\label{TW_rot}
    \delta E^a_{~i} &= 
    \frac{1}{1+ \alpha}\Big[\zeta^a p_b - p^a \zeta_b \Big]E^b_{~i} = \Theta^a_{~b}(\zeta, p) E^b_{~i}
\end{align}
implying that it is not enough for $p^a$ to transform under infinitesimal boosts, but the spatial vielbein $E^a_{~i}$ need to transform with an infinitesimal $\zeta^a$ and $p^a$ dependent rotation parametrised by $\Theta_{ab}(\zeta, p)$.

The transformation \eqref{delta_p} can easily be generated by the function $\alpha \zeta^a \pi_a$,
\begin{align}
    \{p^a, \alpha \zeta^b \pi_b\} = \alpha \zeta^a,
\end{align}
while the vielbein rotation \eqref{TW_rot} can be achieved by adding a multiple of $\Jcal^{ab}= \pi^{i[a}E^{b]}_{~i}$,
\begin{subequations}
\begin{align}
    \Kbb^a &= \alpha \pi^a + \frac{2}{1+ \alpha}\Jcal^{a b}p_b\\
\label{K_def}
    \Kbb[\vec{\zeta}] &= \Dint x \Big[ \alpha \zeta_a \pi^a + \Theta_{ab}(\zeta,p)\Jcal^{ab}], \qquad \Theta_{ab} = \frac{2}{1+\alpha }\zeta_{[a}p_{b]}
\end{align}
\end{subequations}
where $\Theta_{ab}$ corresponds to the infinitesimal Thomas–Wigner rotation, a result of two non-collinear boosts.\footnote{Note that this differs from \cite{Faraji:2024enq} where their action has a different boundary term in the action. Their action is of the form $(\partial e)^2$ \cite{Peldan:1993hi}, which breaks the Lorentz invariance of the action up to a boundary term, producing explicit $\dot{p}^a$ dependence and non-trivial primary constraints.}

With this generator, we can now conclude that,
\begin{align}
    \{e^A_{~\mu}(x), \Kbb[\vec{\zeta}]\} = \theta^A_{~B}(x)e^B_{~\mu}(x), && \theta_{ab}=0, \quad \theta_{a0}= \zeta_a,
\end{align}
generating the correct transformation. A general Lorentz transformation can thus be generated by $\Jbb[\vartheta] + \Kbb[\vec{\zeta}]$,
\begin{align}
    \{e^A_{~\mu}(x), \Jbb[\vartheta] + \Kbb[\vec{\zeta}] \} = \theta^A_{~B}(x) e^B_{~\mu}(x), && \theta^A_{~B} = 
    \begin{pmatrix}
    0 & \zeta_b \\
    \zeta^a & \vartheta^a_{~b}
    \end{pmatrix}.
\end{align}
As $\pi^i_{~a}$ transforms only with a rotation under boosts,
\begin{align}
    \{\pi^i_{~a}(x), \Kbb[ \zeta]\}= - \pi^i_{~b}(x)\Theta^b_{~a}(x),
\end{align}
SO($D$) invariant terms like $\pi^i_a E^a_{~j}$ remain invariant under the induced boost as well, from which it directly follows that the Hamiltonian constraint is invariant,
\begin{align}
    \{\Jbb[\vartheta], \Rcal[\chi]\}=0, && \{\Kbb[\vec{\zeta}], \Rcal[\chi]\}=0.
\end{align}
We previously concluded that $E^a_{~i}, \pi^i_{~a}, p^a,$ and $\pi_a$ all transform tensorially under the action of $\Jbb[\vartheta]$, implying,
\begin{align}
    \{\Jbb[\vartheta], \Kbb^a\} = -\vartheta^a_{~b}\Kbb^b, \qquad \Longrightarrow \qquad \{\Jbb[\vartheta], \Kbb[\vec{\zeta}] \}=  \Kbb[\vartheta\cdot \vec{\zeta}], \qquad (\vartheta\cdot \vec{\zeta})_a = \vartheta_{ab}\zeta^b.
\end{align}
When we evaluate the action of $\Kbb[\vec{\zeta}]$ on $\pi_a$, we obtain,
\begin{align}
\label{delta_pi}
    \{\pi_a, \Kbb[\vec{\zeta}] \} = - \frac{\zeta^b \pi_b}{\alpha}p_a + \{\pi_a, \Theta_{bc} \}\Jcal^{bc},
\end{align}
which only generates the correct transformation $\delta \pi_{~a}=-\zeta^b \pi_b p_a/\alpha$ on the primary constraint surface, $\Jcal^{ab} = 0$. This means that the final bracket in the Lorentz algebra only holds weakly,
\begin{align}
    \{\Kbb[\vec{\zeta}],\Kbb[\vec{\zeta}'] \} \approx \Jbb[\vec{\zeta}\wedge \vec{\zeta}'], \qquad (\vec{\zeta}\wedge \vec{\zeta}' )_{ab} = \zeta_a \zeta'_b - \zeta_b \zeta'_a.
\end{align}
One could imagine trying to "cure" the additional term in \eqref{delta_pi} by modifying $\Kbb^a$. But if the modified boost generator $\widehat{\Kbb}^a$ is not to change $\delta E^a_{~i}$, $\delta \pi^i_{~a}$ or $\delta p^a$, any additional piece can only be a function of $p^a$, and $\Kbb^a+f(p^2) p^a$ bracketed with $\pi_b$ can never produce terms proportional to $\Jcal^{ab}$ to cancel $\{\pi_a, \Theta_{bc}\}\Jcal^{bc}$. 

The full SO$(1,D)$ algebra then reads,
\begin{subequations}
\begin{align}
     \{\Jbb[\vartheta], \Jbb[\vartheta']\} &= \Jbb\big[[\vartheta, \vartheta']\big],\\
     \{\Jbb[\vartheta], \Kbb[\vec{\zeta}] \}&=  \Kbb[\vartheta\cdot \vec{\zeta}],\\
     \{\Kbb[\vec{\zeta}],\Kbb[\vec{\zeta}'] \} &\approx \Jbb[\vec{\zeta}\wedge \vec{\zeta}'].
\end{align}
\end{subequations}
To expand this to the full Lorentz covariant diffeomorphism algebra, the spatial diffeomorphism generator must be modified to also transform $p^a$ and $\pi_a$. This is easily done by including a term $p^a \Lcal_{\vec{\xi}}\pi_a $,
\begin{align}
    \widehat{\Rcal}[\vec{\xi}] = \Dint x\, \Big[\xi^i\Rcal_i + p^a \Lcal_{\vec{\xi}} \pi_a \Big],
\end{align}
so that,
\begin{align}
    \{p^a, \widehat{\Rcal}[\vec{\xi}] \} = - \Lcal_{\vec{\xi}} p^a, && \{\pi_a, \widehat{\Rcal}[\vec{\xi}] \} = - \Lcal_{\vec{\xi}} \pi_a
\end{align}
from which one can then compute,
\begin{align}
    \{\Jbb[\vartheta], \widetilde{\Rcal}[\vec{\xi}]\}= \Jbb[\Lcal_{\vec{\xi}}\vartheta], && \{\Kbb[\vec{\zeta}], \widehat{\Rcal}[\vec{\xi}]\}= \Kbb[\Lcal_{\vec{\xi}}\zeta],
\end{align}
providing the final brackets in the rotational-boost decomposition of the Lorentz covariant diffeomorphism algebra.

\section{Summary}
This note has provided a detailed and self-contained derivation of the Hamiltonian formulation of general relativity in terms of vielbein variables in $d=D+1$ dimensions. Starting from a $D+1$ foliation of spacetime and an adapted decomposition of the vielbein, we derived both a manifestly Lorentz-covariant and an $\mathrm{SO}(D)$-covariant phase-space action and Hamiltonian. We identified the primary constraints associated with non-dynamical Lorentz fields, and obtained the Hamiltonian and momentum constraints. We showed how the spatial diffeomorphism generator had to be modified by terms proportional to the Lorentz generator in order for it to act correctly on the full vielbein phase space, and we verified that the complete set of constraint brackets closed into the expected first-class, Lorentz-covariant diffeomorphism algebra. Additionally, we related the vielbein formulation to the standard metric phase-space action, clarified how the different constraint representatives agree on the primary constraint surface while differing in their action on the internal Lorentz degrees of freedom, and explained why their vielbein form was important for the constraint algebra. We showed how the Lorentz covariant Lorentz generator could be decomposed into rotations and boosts, at the expense of making the algebra close only weakly. Finally, we recovered the full local Lorentz symmetry of the $\mathrm{SO}(D)$-covariant theory by constructing the boost and modified generators, and computed their first-class algebra.

\section*{Acknowledgements}
We would like to thank Fawad Hassan for his insight and discussions on the topic.


\paragraph{Funding information}
DB is supported by Carl Tryggers foundation, CTS 24:3698.


\begin{appendix}
\numberwithin{equation}{section}

\section{\texorpdfstring{$\Jcal^{AB}$}{JAB} as Lorentz Momenta}
\label{sec:Lorentz_momenta}
In this appendix, we demonstrate that $\Jcal^{AB}$ corresponds to the momenta conjugate to the Lorentz degrees of freedom.

A convenient way to isolate the local Lorentz degrees of freedom in $E^A_{~i}$ is to factor out an explicit Lorentz matrix,
\begin{align}
\label{E_Lambda_Ebar}
    E^A_{~ i} = \Lambda^A_{~ B}\bar E^B_{~ i},
\end{align}
where $\bar E^A_{~i}$ is a gauge-fixed representative which only depends on the degrees of freedom of the spatial metric $\gamma_{ij}$. The Lorentz matrix $\Lambda^A_{~ B}$ is a $d\times d$ matrix, but is parametrised by only $d(d-1)/2$ degrees of freedom, i.e. $D=d-1$ boost parameters and $D(D-1)/2$ rotational angles. These are conveniently parametrised in terms of the Cayley transform,
\begin{align}
\label{Cayley}
    \Lambda^A_{~ B} = \big[(\eta + w)^{-1}\big]^{AC}[\eta- w]_{CB} && \Longleftrightarrow && w_{\!AB}=\eta^{}_{AD}\big[(\id+\Lambda)^{-1}\big]^{\!D}_{~C}\big[\id-\Lambda \big]^{\!C}_{~B},
\end{align}
where $w_{AB}=-w_{BA}$ parametrises the Lorentz fields.

If we now consider the momenta conjugate to $w_{AB}$, defined by the action \eqref{EH_decomp} and use \eqref{E_Lambda_Ebar}, we get,
\begin{align}
    \frac{\delta\Scal}{\delta \dot{w}_{AB}} = \frac{\delta\Scal}{\delta \dot{E}^C_{~ i}}
    \frac{\delta\dot{E}^C_{~ i}}{\delta \dot{w}_{AB}} = \pi^i_{~ C}\frac{\delta \dot{\Lambda}^C_{~ D}}{\delta \dot{w}_{AB}}(\Lambda^{-1})^D_{~ E}E^E_{~ i}.
\end{align}
Using \eqref{Cayley} one finds,
\begin{align}
     \frac{\delta \dot{\Lambda}^C_{~ D}}{\delta \dot{w}_{AB}}= -\tfrac{1}{2}\big(\eta^{C[A}+ \Lambda^{C[A} \big)\big(\delta^{B]}_D + \Lambda^{B]}_{~ D} \big),
\end{align}
which implies,
\begin{align}
    \frac{\delta\Scal}{\delta \dot{w}_{AB}}=-\frac12(\id+\Lambda^{-1})^{A}{}_{C}(\id+\Lambda^{-1})^{B}{}_{D}\Jcal^{CD}.
\end{align}
This shows that the momenta conjugate to the Lorentz degrees of freedom $w_{AB}$ are proportional to the primary constraints $\Jcal^{CD}$. 

\section{Velocity Inversion and the Pseudoinverse}
\label{pseudo_inv_method}

In this appendix we will provide an alternative method for inverting the velocity-momentum relation using the Moore–Penrose pseudoinverse.

The definition of the momentum conjugate to $E^A_{~i}$ \eqref{pi} can be put in the form,
\begin{align}
\label{pi=MEdot}
    \pi^i_A= M^{ij}_{AB}\dot{E}^B_{~j},
\end{align}
where $M^{ij}_{AB}$ is the symmetric $\big((A,i) \leftrightarrow (B,j)\big)$ matrix,
\begin{align}
    M^{ij}_{AB} &=  \mpl \frac{\sqrt{\gamma}}{N}\Big[E^i_{~B}E^j_{~A}-2E^i_{~A}E^j_{~B} + \gamma^{ij}P_{AB}\Big].
\end{align}
Since we want to solve this relation for the velocities $\dot{E}^A_{~i}(\pi)$, we need to invert the matrix $M$. However, $M$ is singular and has a non-trivial kernel, so that for arbitrary coefficients $\lambda_i$ and $\lambda_{AB}=-\lambda_{BA}$,
\begin{align}
    M_{AB}^{ij}\lambda_j X^B = 0, && M_{AB}^{ij}\lambda^B_{~C}E^C_{~j}=0.
\end{align}
However, the general solution of \eqref{pi=MEdot} for $\dot{E}^A_{~i}$ can be constructed using the Moore–Penrose pseudoinverse. A generalised inverse $M^+$ of a symmetric matrix has the property,
\begin{align}
\label{gen_inv}
    M^+ M= \Pi,
\end{align}
where $\Pi$ is the projector onto the image of $M$, $\mathrm{Im}(M)=\{Y^i_A | Y^i_A = M^{ij}{}_{\!AB}Z^B_j\}$.\footnote{In general $\Pi$ projects onto the image of $M^{\T}$, but since $M$ is symmetric, $\mathrm{Im}(M^{\T})=\mathrm{Im}(M)$ } However, this does not uniquely determine $M^+$, but if we also impose,\footnote{In general the transpose is generalised to the adjoint $\dagger$ induced by a positive definite inner product, which in our case is given by $ \langle U, V\rangle =P_{AB}\gamma^{ij}U^A_{~i} V^B_{~j},$ for vectors $U^A_{~i}$ and $V^A_{~i}$. }
\begin{align}
\label{MP_inv}
    MM^+M = M, && M^+MM^+=M^+, && (MM^+)^{\T} = MM^+, && (M^+M)^{\T} = M^+M,
\end{align} 
the generalised inverse uniquely becomes the Moore–Penrose pseudoinverse.

The projector $\Pi$ onto the image of $M$ takes the form,
\begin{align}
    \Pi^{Ai}_{Bj}= \tfrac{1}{2}\Big[P^A_B \delta^i_j+ E^{Ai}E_{Bj} \Big]= E^{Ak}E_{B(k}\delta^i_{j)},
\end{align}
and by making an Ansatz using the available canonical objects, one can use \eqref{gen_inv} and \eqref{MP_inv} to derive the pseudoinverse,
\begin{align}
    (M^+)_{ij}^{AB} = \frac{N}{4\mpl \sqrt{\gamma}}\Big[ \gamma_{ij}P^{AB}+ E^A_{~j}E^B_{~i}-\tfrac{2}{d-2}E^A_{~i}E^B_{~j}\Big].
\end{align}
Since multiplying \eqref{pi=MEdot} by $M^+$ only yields the projection onto the image of $M$, $\Pi \dot{E}$, we need to add arbitrary combinations of the kernel $\lambda_i X^A, \, \lambda^A_{~B}E^B_{~i}$ to obtain the general solution,\footnote{This is the origin of the Dirac procedure of adding the primary constraints to the canonical Hamiltonian and constructing the total Hamiltonian.}
\begin{align}
\label{Edot_sol}
    \dot{E}^A_{~i} = (M^+)^{AB}_{ij}\pi^j_B+ \lambda_i X^A + \lambda^A_{~B}E^B_{~i}.
\end{align}
This can now be used, for example, in the Legendre transform to construct the phase-space action or the Hamiltonian. E.g. the canonical one-form is directly given by the direct contraction of \eqref{Edot_sol} with $\pi^i_A$,
\begin{align}
    \pi^i_A\dot{E}^A_{~i} &= \pi^i_A(M^+)^{AB}_{ij}\pi^j_B+ \lambda_i X^A\pi^i_A + \lambda^A_{~B}\pi^i_AE^B_{~i}\notag\\
    &=\pi^i_A(M^+)^{AB}_{ij}\pi^j_B + \lambda'_{AB}\Jcal^{AB},
\end{align}
where we, in the second line, have used that $\Jcal^{AB}= \pi^{i[A}E^{B]}_{~i}$ and that since $\lambda_{AB}$ is arbitrary we can write $\lambda'_{AB}= \lambda_{AB}+ 2X_{[A}\lambda_{B]}$, where $\lambda_B = \lambda_i E^i_{~B}$.

\section{Metric Action and Modified Generators}
\label{sec:metric_action}

In Section \ref{sec:Phase_space_Action} and \ref{sec:so_phase_space_action} we derived the Lorentz and SO(\textit{D}) covariant phase-space actions. Here, we comment on the apparent differences from the more common metric formulation, where the phase-space action reads,
\begin{subequations}
\begin{align}
\label{metric_action}
    \Scal^\gamma&=\dint x\Big[\pi^{ij}\dot{\gamma}_{ij}+N\Rcal^\gamma+N^i \Rcal_i^{\gamma}\Big],\\
\label{met_ham}
    \Rcal^\gamma &= \mpl\sqrt{\gamma}\rD\! R + \tfrac{1}{\mpl \sqrt{\gamma}}\Big[\tfrac{1}{d-2}\big(\pi^i_{~ i}\big)^2 - \pi_{ij}\pi^{ij} \Big],\\
    \Rcal^\gamma_i &= 2 \gamma_{ij}\rD\nabla_k \pi^{jk},
\end{align}
\end{subequations}
with the momenta conjugate to the spatial metric given by,
\begin{align}
    \pi^{ij}= \frac{\delta \Scal}{\delta \dot{\gamma}_{ij}}=\tfrac{1}{2}\pi^{(i}_{~A} E^{j)A}_{\phantom{A}}.
\end{align}
Here and below, we will use the covariant notation, but obtain the SO($D$)-covariant expressions if we set $X^A=(1,0)$, $E^0_{~i} =\pi^i_{~0}=0$ and restrict $A,B, \dots$ to $a,b, \dots$.

If we compare the canonical one-forms $\pi^{ij}\dot{\gamma}_{ij}$  and $\pi^i_A\dot{E}^A_{~ i}$, we can, from the identity,
\begin{align}
    2E_{Ai}\dot{E}^A_{~j}&= \dot{\gamma}_{ij}+ 2E_{A[i}\dot{E}^A_{~j]},  
\end{align}
derive the relation,
\begin{align}
\label{one_form}
    \pi^{ij}\dot{\gamma}_{ij}= \pi^i_A\dot{E}^A_{~ i} + \Big[E^i_{[A}\dot{E}^{}_{B] i}-X_{[A}\dot{X}_{B]}\Big]\Jcal^{AB},
\end{align}
where we have used $X_A\dot{E}^A_{~i}=-E_{Ai}\dot{X}^A$. The bracketed expression is the generalised velocity of the Lorentz degrees of freedom, conjugate to  $\Jcal^{AB}$. These are precisely the first two terms in \eqref{Omega_def},
\begin{align}
\label{omegaJ}
    W_{AB}\Jcal^{AB}&=\Big[E^i_{[A}\dot{E}^{}_{B] i}-X_{[A}\dot{X}_{B]}- E^i_{[A}E^j_{B]}\rD\nabla_i N_j -\tfrac{N}{2}E^i_{[A}E^j_{B]}\Kcal_{ij}\Big]\Jcal^{AB},
\end{align}
which is part of the additional term in the vielbein actions \eqref{can_action} and \eqref{EH_phase_space_soD}. The additional terms come from the difference in the Hamiltonian and momentum constraints in the metric and vielbein formulations. 

Let's consider the metric form of the Hamiltonian constraint and substitute $\gamma_{ij}=E^A_{~ i}E_{Aj}$ and $\pi^{ij}=\tfrac{1}{2}\pi^{(i}_{~A}E^{j)A}_{\phantom{A}}$ into $\Rcal^\gamma$ \eqref{met_ham}, 
\begin{align}
\label{metric_ham}
    \Rcal^\gamma &= \mpl \sqrt{\gamma}\rD \!R+ \tfrac{1}{4\mpl \sqrt{\gamma}}\Big[\tfrac{1}{d-2}\big(\pi^i_AE^A_{~ i}\big)^2-E^{A(i}\pi^{j)}_{ A}E^B_{~j}\pi_{i B}\Big].
\end{align}
This differs from the vielbein expressions \eqref{vielbein_ham} and \eqref{R_constraint_soD} by an anti-symmetric part,
\begin{align}
    \Rcal &= \mpl \sqrt{\gamma}\rD \!R+ \tfrac{1}{4\mpl \sqrt{\gamma}}\Big[\tfrac{1}{d-2}\big(\pi^i_AE^A_{~ i}\big)^2-E^A_{~ j}\pi^j_{~ B}E^{B}_{~ i}\pi^i_{ A}\Big]\notag\\
\label{new_ham}
    &= \Rcal^\gamma +\tfrac{1}{2}E^i_{[A}E^j_{B]}\Kcal_{ij}\Jcal^{AB},
\end{align}
where we have used $E^{A[i}_{}\pi^{j]}_{A} = E^i_{A}E^j_B\Jcal^{AB}$ and the expression for $\Kcal_{ij}$ \eqref{Kij}. So $\Rcal$ and $\Rcal^\gamma$ differ only on the primary constraint surface and the additional term produces the last term in $W_{AB}\Jcal^{AB}$ \eqref{omegaJ}.

Note that, since $\Rcal^\gamma$ does not depend on the anti-symmetric combination of $\pi^{i}_{~A}E^{jA}$, it does not act as a generator of temporal diffeomorphisms on all components of $E^A_{~i}$ and $\pi^i_A$. For example, the anti-symmetric combination,
\begin{align}
    E^{\phantom{A}}_{A[i}\{E^A_{~j]},\Rcal^{\gamma}[\chi]\} = \gamma_{k[i}\frac{\delta \Rcal^\gamma[\chi]}{\pi_{j]k}}=0,
\end{align}
vanishes identically, implying that it does not generate any evolution of the Lorentz fields. While,
\begin{align}
    E^{\phantom{A}}_{A[i}\{E^A_{~j]},\Rcal[\chi]\}= \frac{\chi}{2\mpl \sqrt{\gamma}}E^A_{~[i}E^B_{~j]}\Jcal_{AB},
\end{align}
rotate the Lorentz degrees of freedom through the normal deformations of the spatial hypersurface generated by a temporal diffeomorphism.

The term $E^i_{[A}E^j_{B]}\rD\nabla_i N_j$ in \eqref{omegaJ} comes from the difference in the momentum constraints in the metric and vielbein formulation, $\Rcal_i^\gamma= 2\gamma_{ij}\rD\nabla_k \pi^{jk}$ and $\widetilde{\Rcal}_i= E^A_{~ i} \rD \Dcal_j \pi^j_A$. Substituting $\gamma_{ij}=E^A_{~ i}E_{Aj}$ and $\pi^{ij}=\tfrac{1}{2}\pi^{(i}_{~A}E^{j)A}_{\phantom{A}}$ into 
$\Rcal^\gamma_i$, yields,
\begin{align}
\label{metric_mom_const}
    \Rcal^\gamma_i = 2\rD\nabla_j \pi^j_{~ i}= 2\rD \Dcal_j \pi^j_{~ i} = E^A_{~(i}\rD\Dcal^j \pi^{}_{j)A},
\end{align}
so that the vielbein expression $\widetilde{\Rcal}_i=E^A_{~ i}\rD\Dcal_j \pi^j_A$ can be written as,
\begin{align}
    \widetilde{\Rcal}_i =E^A_{~ (i}\rD\Dcal^j \pi_{j)A}+E^A_{~ [i}\rD\Dcal^j \pi_{j]A}^{\phantom{A}} = \Rcal^\gamma_i + \rD \Dcal(E_{Ai}E_{Bj}\Jcal^{AB}),
\end{align}
where we in the last step have used that $E^A_{~[i}\pi_{j]A}^{\phantom{A}}= E_{Ai}E_{Bj}\Jcal^{AB}$ and the vielbein compatibility of $\rD\Dcal_j$. If we multiply by $N^i$ and integrate, we get, modulo a boundary term,
\begin{align}
\label{momentum_const}
   \Dint x\sqrt{\gamma} N^i  \Rcal^{\gamma}_i = \Dint x\sqrt{\gamma} \Big[N^i\widetilde{\Rcal}_i -E^i_{[A}E^j_{B]}\rD\nabla_i N_j \Jcal^{AB}\Big],
\end{align}
in agreement with the third term in \eqref{omegaJ}. 

The momentum constraint $\Rcal^\gamma_i$ has the similar problem as the Hamiltonian one $\Rcal^\gamma$, i.e. it does not generate any transformation on the Lorentz fields along a spatial diffeomorphism. So while,
\begin{align}
    \{ \gamma_{ij}, \Rcal^\gamma[\vec{\xi}]\} = -\Lcal_{\vec{\xi}}\gamma_{ij} \qquad \Longrightarrow \qquad E_{A(i}\{E^A_{~j)}, \Rcal^\gamma[\vec{\xi}]\} = -E_{A(i}\Lcal_{\vec{\xi}} E^A_{~j)}
\end{align}
the anti-symmetric part vanishes trivially,
\begin{align}
    E_{A[i}\{E^A_{~j]}, \Rcal^\gamma[\vec{\xi}]\} = 0,
\end{align}
thus not providing the correct transformation for the Lorentz degrees of freedom contained in $E^A_{~i}$.

In Appendix \ref{sec:mod_gen_1} we show that also $\widetilde{\Rcal}_i$ needs to be modified $\widetilde{\Rcal}_i \to \Rcal_i$ \eqref{mod_dif_gen} to generate pure diffeomorphisms on $E^A_{~i}$ and $\pi^i_A$. For example, using \eqref{mod_dif_gen}, which again differs only from $\Rcal^\gamma_i$ up to terms proportional to $\Jcal^{AB}$, the vielbein transforms as,
\begin{align}
    \{E^A_{~i},\vec{\Rcal}[\vec{\xi}]\}&= - \Lcal_{\vec{\xi}}E^A_{~i} =-\Big[\tfrac{1}{2}E^{Aj}\Lcal_{\xi}\gamma_{ji} + \theta^A_{~B}E^B_{~i}\Big],\\
    \theta_{AB}&= E^{i}_{~[A}\Lcal_{\xi}E_{B]i}
\end{align}
which is the expected transformation behaviour which additionally can be decomposed into a Lie drag of the metric variables and a field dependent rotation $\theta$ of the Lorentz  degrees of freedom.

We stress that the additional terms in $\Rcal$ and $\Rcal_i$ are proportional to $\Jcal^{AB}$, thus they only differ from the metric constraints $\Rcal^\gamma$ and $\Rcal^\gamma_i$ weakly. This means that the dynamics will remain correct even if the action is written in terms of the metric expressions $\Rcal^\gamma$ and $\Rcal^\gamma_i$, but the diffeomorphism algebra would only close on the primary constraint surface in contrast to the diagonalised form \eqref{full_alg}. 

\section{Spatial Diffeomorphism Generator}
\label{sec:mod_gen_1}
In this appendix, we demonstrate how to construct the generator of spatial diffeomorphisms in the Lorentz and SO(\textit{D}) covariant formulations. While we will use the notation for the covariant formalism, the SO(\textit{D}) covariant expressions can be obtained by substituting $X^A=(1,0)$, $E^0_i =\pi^i_0=0$ and restrict $A,B, \dots$ to $a,b, \dots$.

Recall that an infinitesimal spatial diffeomorphism, generated by a vector field $\xi^i(x)$, acts on $E^A_{~i}$ (a spatial covector) and $\pi^i_{A}$ (a spatial vector density) as,
\begin{subequations}
\label{diff}
\begin{align}
\label{diff_vielbein}
    \delta_{\vec{\xi}}\,E^A_{~ i} 
    &= \Lcal_{\vec{\xi}}\,E^A_{~ i} 
    = E^A_{~ j}\partial_i \xi^j+\xi^j \partial_j E^A_{~ i},\\
\label{diff_momenta}
    \delta_{\vec{\xi}}\,\pi^i_A 
    &= \Lcal_{\vec{\xi}}\,\pi^i_A 
    = \xi^j\partial_j \pi^i_A-\pi^{j}_{~ A}\partial_j \xi^i + \pi^{i}_{~ A}\partial_j \xi^j.
\end{align}
\end{subequations}
It is easy to construct a generator that reproduces these transformations through the Poisson bracket,
\begin{align}
\label{diff_gen}
    \vec{\Rcal}[\vec{\xi}] 
    =\Dint x\, \xi^i\Rcal_i
    =
    \Dint x\, E^A_{~ i}\Lcal_{\vec{\xi}}\,\pi^i_A
    =-\Dint x\, \pi^i_A\Lcal_{\vec{\xi}}\, E^A_{~ i},
\end{align}
which by a direct application of the canonical brackets yields,
\begin{align}
\label{vielbein_diff}
    \{E^A_{~ i}(x), \vec{\Rcal}[\vec{\xi}]\}
    &= -\Lcal_{\vec{\xi}}\, E^A_{~ i}(x), &
    \{\pi^i_A(x), \vec{\Rcal}[\vec{\xi}]\}
    &= -\Lcal_{\vec{\xi}}\, \pi^i_A(x).
\end{align}
Thus $\Rcal_i$ generates pure spatial diffeomorphisms on the canonical vielbein variables.

We can now relate the generator $\Rcal_i$ to $\widetilde{\Rcal}_i= E^A_{~ i}\rD \Dcal_j \pi^j_A$ which appears naturally in the actions \eqref{can_action} and \eqref{EH_phase_space_soD}. The key observation is that on a field with internal Lorentz indices, the Lie derivative differs from a local Lorentz-covariant derivative by a field dependent Lorentz transformation. Concretely, if we add $\xi^j \,\rD\!\omega_{j}{}^A_{~B} E^B_{~ i}$ to \eqref{diff_vielbein}, the combination can be expressed as a vielbein-compatible covariant derivative,
\begin{align}
    \Lcal_{\vec{\xi}} E^A_{~ i} + \xi^j \,\rD\!\omega_{j}{}^A_{~B} E^B_{~ i}
    = \rD\Dcal_i(E^A_{~ j} \xi^j),
\end{align}
where $\rD\!\omega_{iAB}$ is the $D$-spin-connection \eqref{wiAB_D}. Using this identity in \eqref{diff_gen}, we obtain,
\begin{align}
    \Dint x\, \xi^i\Rcal_i 
    &=
    - \Dint x\; \pi^i_A
    \Big[\rD \Dcal_i(E^A_{~ j} \xi^j)-\xi^j \,\rD\!\omega_{j}{}^A_{~B} E^B_{~ i}\Big]
    =
    \Dint x \,\xi^j\Big[\widetilde{\Rcal}_j+ \,\rD\!\omega_{jAB}\Jcal^{AB} \Big].
\end{align}
This implies that $\widetilde{\Rcal}_i$ generates a spatial diffeomorphism together with an internal Lorentz rotation, which the additional term $\rD\!\omega_{iAB}\Jcal^{AB}$ compensates for. We may therefore define a new momentum constraint that generates pure spatial diffeomorphisms,
\begin{align}
\label{mod_dif_gen}
    \Rcal_i &= \widetilde{\Rcal}_i + \rD\!\omega_{iAB}\Jcal^{AB}.
\end{align}
Note that $\Rcal_i$ and $\widetilde{\Rcal}_i$ agree when $\Jcal^{AB}=0$, so $\widetilde{\Rcal}_i$ generates spatial diffeomorphisms on the primary constraint surface, but not the general vielbein variables. 

\section{\texorpdfstring{$D+1$}{D+1} Christoffel and Spin Connection}
\label{sec:Christoffel}

In this appendix we provide the $D+1$ decomposition of Christoffel symbols and spin connection in terms of the lapse, shift, extrinsic curvature and spatial covariant derivative $\rD \nabla_i$.

The Christoffel symbols read,
\begin{subequations}
\begin{align}
\label{Gamma_0ij}
    \Gamma^0_{~ ij}&= \frac{1}{N}K_{ij},\\
\label{Gamma_000}
    \Gamma^0_{~ 0 0}&= \frac{1}{N}\Big[\dot{N} + N^i \partial_i N + N^i N^j K_{ij} \Big] = \frac{\dot{N}}{N}+ N^i\Gamma^0_{~ i0}\\
\label{Gamma_0i0}
    \Gamma^0_{~ i 0}&= \frac{1}{N}\Big[\partial_i N + N^j K_{ij}\Big]=\frac{1}{N}\partial_i N +N^j\Gamma^0_{~ij},\\
\label{Gamma_ij0}
    \Gamma^i_{~ j0}& = \rD\nabla_j N^i-\frac{1}{N}N^i \partial_j N+ N\Big[\gamma^{ik}-\frac{1}{N^2}N^i N^k \Big]K_{kj}= \rD\nabla_j N^i +NK^i_{~ j}-N^i \Gamma^0_{~ 0j},
    \\
\label{Gamma_i00}
    \Gamma^i_{~ 00}&= \dot N^i+ N \partial^i N+ N^k \rD\nabla_k N^i+ 2N K^i_{~k} N^k- N^i \Gamma^0_{~ 00},\\
\label{Gamma_ijk}
    \Gamma^i_{~ jk}&=\rD\Gamma^i_{~ jk} - \frac{1}{N}N^i K_{jk} =\rD\Gamma^i_{~ jk}-N^i \Gamma^0_{~jk},
\end{align}
\end{subequations}
where $\rD \Gamma^{i}_{~jk}$ is the Christoffel symbols of $\rD \nabla_i$,
\begin{align}
    \rD \Gamma^{i}_{~jk} = \tfrac{1}{2}\gamma^{il}\Big[\partial_j \gamma_{lk}+ \partial_k \gamma_{jl}- \partial_l \gamma_{jk}  \Big].
\end{align}
The Lorentz-covariant spin-connection is defined by,
\begin{align}
    \omega_{\mu AB}= e^\nu_{~[A}\rd \nabla_\mu e^{~}_{B] \nu}
\end{align}
and can be decomposed as,
\begin{subequations}
\begin{align}
\label{w0AB}
    \rd\!\omega_{0AB}&=E^{i}{}_{[A}\dot E_{B]i}-X_{[A}\dot X_{B]}-E^{i}{}_{[A}E^j{}_{B]}\rD\nabla_i N_j+2X_{[A}E^{i}\!{}_{B]}\big[\partial_i N+K_{ij}N^{j}\big],\\
    \rd\!\omega_{iAB}&= \rD\!\omega_{iAB}-X_{[A}\partial_i X_{B]}+2X_{[A}E^j\!{}_{B]}K_{ij},\\
\label{wiAB_D}
    \rD\!\omega_{iAB}&=E^j_{[A}\rD\nabla_i E_{B]j}.
\end{align}
\end{subequations}
The spin-connection used in Section \ref{sec:soD_covariant} can easily be obtained by choosing the time gauge in the above, i.e. inserting $X^A = (1,0)$ and setting $E^0_{~ i}=0$, yielding,
\begin{subequations}
\label{time_gauge_w}
\begin{align}
    \rd \!\omega_{0ab}&= E^i_{~[a}\dot{E}_{b]i} -E^i_{~[a}E^j_{~b]}\rD \nabla_i N_j,\\
    \rd \!\omega_{0a0}&= E^a_{~ i}\big[\partial_i N+ K_{ij}N^j\big],\\
    \rd \!\omega_{ia0}&=K_{ij}E^j_{~ a}, \\
    \rd\!\omega_{iab}&=\rD\!\omega_{iab} = E^j{}_{[a} \rD \nabla_i E_{b]j}.
\end{align}
\end{subequations}

\end{appendix}





\bibliography{ref.bib}


\end{document}